\documentclass[11pt]{article}

\usepackage{amsfonts}
\usepackage{amsbsy} 
\usepackage{epsfig}
\usepackage{latexsym}
\input amssym.def
\input amssym.tex

\def\bequ{\begin{equation}}
\def\eequ{\end{equation}}
\def\barr{\begin{array}}
\def\earr{\end{array}}

\def\ben{\begin{equation}}
\def\een{\end{equation}}
\def\bena{\begin{eqnarray}}
\def\eena{\end{eqnarray}}

\newcommand{\sect}[1]{\setcounter{equation}{0}\section{#1}}

\advance \topmargin by -\headheight
\advance \topmargin by -\headsep
\evensidemargin \oddsidemargin
\advance\hoffset by -3mm  
\advance\voffset by  8mm  

\begin{document}
\hfuzz=100pt
\title{{\Large \bf{Special Properties of Five Dimensional \\BPS Rotating Black Holes}}} 
\author{\\C A R Herdeiro\footnote{E-mail:car26@damtp.cam.ac.uk}
\\
\\
Department of Applied Mathematics and Theoretical Physics,
\\ (D.A.M.T.P.)
\\ Centre for Mathematical Sciences,
\\ University of Cambridge, 
\\ Wilberforce Road,
\\ Cambridge CB3 0WA,
 \\ U.K.}

\date{March 2000}
\maketitle

\centerline{DAMTP-2000-13}

\begin{abstract} 
Supersymmetric, rotating, asymptotically flat black holes with a regular horizon are rare configurations in String Theory. One example is known in five spacetime dimensions, within the toroidal compactification of type IIB string theory. The existence of such special solution is allowed by the existence of a Chern-Simons coupling in the Supergravity theory and by the possibility of imposing a self duality condition on the `rotation 2-form'. We further exemplify the use of such duality condition by finding a new Brinkmann wave solution in $D=6$ simple gravity, possessing Killing spinors. We then explore three peculiar features of the aforementioned black holes: 1) Oxidising to $D=10$ the five dimensional configuration may be interpreted as a system of $D1-D5$ branes with a Brinkmann wave propagating along their worldvolume. Unlike its five dimensional Kaluza-Klein compactification, the universal covering space of this manifold has no causality violations. In other words, causal anomalies can be solved in higher dimensions. From the dual SCFT viewpoint, the causality bound for the compactified spacetime arises as the unitarity bound; 2) The vanishing of the scattering cross section for uncharged scalars and sufficiently high angular momentum of the background is shown still to hold at the level of charged interactions. The same is verified when a non-minimal coupling to the geometry is used. Therefore, the `repulson' behaviour previously found is universal for non accelerated observers; 3) The solutions are shown to have a non-standard gyromagnetic ratio of $g=3$. In contrast, the superpartners of a static, BPS, five dimensional black hole have $g=1$. At the semi-classical level, we find that a Dirac fermion propagating in the rotating hole background has $g=2\pm1$, depending on the spinor direction of the fermion being parallel to Killing or `anti-Killing' spinors. 

\end{abstract}

\sect{Introduction}
One of the most appealing features of String Theory is that it provides a finite quantum theory of gravity which reduces to the well tested General Relativity in the appropriate limit. Quantum gravitational effects become particularly important in at least two physical situations - the early universe and black holes physics - which are therefore laboratories to test the novel stringy physics. But whereas some significative progress has been done in understanding microscopics of black holes, stringy cosmology is still in a very early stage. One important reason for this distinction is supersymmetry. 

Supersymmetry has allowed quantitative mapping between classical gravitational configurations and quantum states of strings. But one pays a price: preservation of supersymmetry requires the initial classical configuration  to be dynamically very simple. In particular, it must contain a causal Killing vector field, in order to possess a Killing spinor. Whereas this is not an obstacle in understanding some features of black holes, since we may study extreme Reissner-Nordstr\"{o}m (RN) type solutions, it becomes a major obstacle in studying cosmology, since there seems to be no easy way to reconcile supersymmetry with an expanding universe.

Even within the domain of black holes one has the natural desire of understanding more complex configurations than the aforementioned ones, in particular to include rotation, which we expect will be present in the black holes we find in Nature. Rotation makes the dynamical properties of a black hole spacetime richer, and therefore `less compatible' with supersymmetry. In fact, regular (on and outside a horizon), rotating, supersymmetric black hole solutions are rare configurations in string theory. Let us consider the Heterotic string theory, compactified on a $T^{10-D}$ torus to $9\ge D \ge 4$ spacetime dimensions. For $9\ge D \ge 6$, the low energy effective field theory  will contain $(36-2D)$ one-form gauge potentials, giving rise to electric charges $Q_a$, $a=1...36-2D$. Of course, these charges have no absolute meaning in themselves; they can be mixed together by the $O(10-D,26-D)$ symmetry of the low energy effective action. In particular, the subgroup $[SO(10-D)\times SO(26-D)]/[SO(9-D)\times SO(25-D)]$ has dimension $(34-2D)$. Therefore, a $9\ge D\ge 6$ dimensional black hole solution classified by mass, the maximal number of angular momentum parameters, $l_i$, $i=1...[(D-1)/2]$=rank of $SO(D-1)$, and two electric charges can be used to generate a solution with all $(36-D)$ charges, by the use of the above subgroup. The solution with two charges is called the \textit{generating} solution. Since the $O(10-D,26-D)$ symmetry does not act on the D dimensional metric, the geometry for the most general electrically charged rotating black hole solution can be seen at the level of the generating solution.

In $D=5$, the (reduced) Neveu-Schwarz field can be Hodge dualized to couple to `0-branes'. It carries a magnetic charge for a black hole. Hence, the generating solution will have 3 charges, and the most general electrically charged black hole 26 electric and 1 magnetic charge. In $D=4$, the one form fields  also carry magnetic charges for black holes. Hence the generating solution has 4 charges and the most general black hole solution 56 charges, 28 electric and 28 magnetic\footnote{The four dimensional case is actually more complex and the generating solution should have five charges; a case which has not been dealt with in the literature \cite{cvey4}.}.

The geometry of these generating solutions in the BPS limit has the following behaviour:
\begin{description}
\item[i)] In $D=4$ \cite{cvey4}; the spacetime with non-vanishing angular momentum yields a naked (timelike) singularity. Putting all charges equal it reduces to the extreme (i.e.  $Q^2=M^2$ and arbitrary $J$)  Kerr-Newman (KN) spacetime of Einstein-Maxwell theory;
\item[ii)] In $D=5$; a regular rotating black hole spacetime \cite{cveyoum}. Putting all charges equal it reduces to the Breckenridge-Myers-Peet-Vafa (BMPV) spacetime \cite{BMPV}, of $N=2$, $D=5$ Supergravity; 
\item[iii)] In $D\ge6$; a naked singularity. When only one $l_i$ is non-zero it reduces to a better behaved null singularity \cite{cvey6}. 

\end{description}
Regularity, therefore, distinguishes five dimensional, BPS rotating black holes. But why?

In section 2 we perform a comparative analysis between the extreme KN black hole and the BMPV black hole. In particular, we emphasize how a Hodge self duality condition, which can be implemented in a four dimensional transverse space to the world volume, allows for the harmonic function in the metric and the Sagnac connection \cite{gibher} to be independent. This is essential for the properties of the solution. We then give a simple example of using such duality condition. Specifically we find a `supersymmetric' Brinkmann wave type solution to $D=6$ pure gravity and mention the corresponding five dimensional Kaluza-Klein rotating black hole.

In section 3 we exhibit the $D=10$ description of five dimensional, BPS, rotating black holes. We also analyse the interesting causal structure of the $D=5$ black holes from the ten dimensional perspective, finding that the unavoidable Closed Timelike Curves (CTC's) of the five dimensional solution acquire a trivial character in ten dimensions (in the language of \cite{carter}). As far as we are aware this is the first example of resolving causal anomalies in higher dimensions. In order to give a complete description of the causal structure of this black holes, we review the string theory dual picture of the compactified spacetime in terms of an $N=4$ Super Conformal Field Theory, which has allowed the microscopic computation of the entropy \cite{BMPV}. The point we wish to stress is the loss of unitarity in this theory when the spacetime undergoes a loss of causality. This unitarity bound can be seen by looking at general unitarity requirements for the conformal weights of SCFT operators. Alternatively one may examine the representations of an $N=2$ SCFT \cite{kent} and the condition for these to be unitary, that arises from analysing the sign of the determinant of the Verma module. 

\bigskip

Spacetimes where causality is violated may have odd effects related to geodesic motion. One such example is the ``totally imprisoned incompleteness'' found for the Lorentzian Taub-NUT metric \cite{hawell}. The phenomenon consists in the existence of a family of null geodesics which are imprisoned inside a compact region of spacetime whose boundary they reach within finite affine length. Such geodesics are therefore incomplete. However, they meet neither an s.p. (divergence of a curvature invariant) nor a p.p. singularity (divergence of the components of the Riemann tensor in a parallelly propagated frame). 

The hypersurface from which they cannot be extended separates a causally well behaved spacetime region (the region initially found by Taub) from the outer reaches where closed timelike curves exist. We should remark, however, that there exists another family of null geodesics that passes through these surfaces. 

In section 4.1 we explore an interesting effect arising for the BMPV spacetime whenever CTC's arise outside the horizon - the over-rotating case - firstly pointed out in \cite{gibher}. The phenomenon -a `repulson' effect- is shown still to be present when we consider test particles with charge, naturally following ``charged geodesics''. For scalar waves we also try a non-minimal coupling to the geometry, in the Klein-Gordon equation, by including the Ricci scalar. Separation of variables can still be achieved, and a radial equation obtained. We conclude the universality of the effect for non accelerated observers. As for the Taub-NUT case we could find neither p.p. nor s.p. singularities, and nothing special seems to occur for an accelerated observer trying to overcome this natural obstacle. We also note that this repulson behaviour is distinct from singular repulsons recently studied and resolved in the context of AdS/CFT \cite{peetpol}.

\bigskip
The uniqueness or `no-hair' theorems for black holes tell us that a small set of independent quantities completely determines black hole spacetimes. Within Einstein-Maxwell theory such set includes solely ADM mass ($M$), angular momentum ($J^i$), electric ($Q$) and magnetic ($P$) charges. In spite of such restriction, black holes may have other multipole moments, albeit not independent. The magnetic dipole moment ($\mu^i$) is the most natural one, since we expect any electrically charged rotating object to have it. In $D=4$ it is related to the previous quantities by $\mu^i$ $\propto Q J^i/(2M)$, therefore defining a constant of proportionality which is called the gyromagnetic ratio. 

In ordinary electrodynamical systems, it follows from the definitions that $g=1$ for any rotating homogeneous distribution of charges. But for four dimensional Kerr-Newman black holes, $g=2$ \cite{carter} in analogy with the quantum mechanical value for the electron (up to loop quantum corrections). The same value for $g$ is found for other heterotic black holes in four dimensions \cite{sen}\footnote{The charges for the Kerr-Newman-Sen \cite{sen} black hole and for the `electron' have different origins, however, in heterotic string theory. Therefore there is a distinction in their gyromagnetic ratios \cite{dufflr}.}, and some p-brane solutions in other dimensions, namely the membrane in $D=11$ \cite{balaktw}. However, this is by no means universal. Kaluza-Klein black holes have a gyromagnetic ratio that depends on the several metric parameters \cite{larsen}. In the limit of large electric charge and vanishing magnetic charge, $g$ approaches one. This is the natural value for massive Kaluza-Klein modes in five dimensional Kaluza-Klein theory \cite{jap}. The classical value arises because the charge is orbital motion in the compact direction. Another Kaluza-Klein black hole is the Dirichlet 0-brane supergravity solution in $D=10$. It is known to have $g=1$ \cite{duffg1}. In fact, it is well known that the identification of the spectrum of IIA D0-branes with the Kaluza-Klein modes of the 11D graviton was a major guideline in the discovery of M-theory \cite{witten}.

In finding the value of $g$ for the $M2$ and $D0$ branes, the superpartners technique was used. This method is usually applied to static BPS solutions, yielding a new solution with fermionic hair.\footnote{The new configuration is only a solution to the full supergravity equations of motion if the method is carried out to all orders. Otherwise it solves the equations of motion only to some order in fermionic parameters.} The interpretation of these ``superpartners'' is ambiguous due to the existence of bilinears of odd Grassmann numbers in the bosonic fields. Nevertheless, the method is useful in reading quantities like the gyromagnetic ratio. The technique was first used in \cite{aiche} and applied to the extreme RN background. The superpartners were found to have $g=2$.

In section 4.2, we compute the gyromagnetic ratio for both the BMPV background and its superpartners showing they are  different. This is in contrast with the four dimensional case, where superpartners of the extreme RN background have the same $g$ as the Kerr-Newman. We then analyse the problem of a Dirac fermion in the BMPV spacetime and show it is isomorphic to the problem of a non-minimally coupled Dirac field in flat space. This coupling alters the usual value for $g$, creating a matching between the behaviour of the elementary particle and either the BMPV black hole or the superpartner of the static $D=5$ RN spacetime.

We close with a discussion.

\sect{BMPV versus Extreme KN and the self-duality condition}

In $D=4,5$ the simplest supergravity theories containing a Maxwell field are the $N=2$ theories. The bosonic truncations of the general theories are, respectively, the Einstein-Maxwell theory and Einstein-Maxwell-Chern-Simons theory, with actions given by 
\bequ
\mathcal{S}^{(4)}=\frac{1}{16\pi G_4}\int d^4x \sqrt{-g} \left[R-\frac{1}{4}F^2\right], \ \ \  \mathcal{S}^{(5)}=\frac{1}{16\pi G_5}\int d^5x \left[ \sqrt{-g}(R-F^2)-\frac{2}{3\sqrt{3}} A\wedge F\wedge F\right].
\eequ

Finding supersymmetric solutions to these theories involves solving the gravitino variation equations which are, respectively,

\bequ  
\displaystyle{ D\epsilon  -\frac{1}{4}F_{a b}\Gamma^{ab}\Gamma\epsilon=0, \ \ \     D\epsilon   +\frac{i}{4\sqrt{3}}\left(e^a\Gamma^{bc}_{\ \ \ a}-4e^b\Gamma^c\right)F_{bc}\epsilon=0,} 
\eequ
where we have denoted the covariant derivative acting on spinors by $D\epsilon=d\epsilon +1/4w_{ab}\Gamma^{ab}\epsilon$. For the four dimensional theory all solutions possessing Killing spinors were found by Tod \cite{tod1}. They split into two families, according to the existence of a null or a timelike Killing vector field. The former type corresponds to a family of gravitational waves, carrying electromagnetic fields, with the standard plane-fronted waves with parallel rays (pp-waves) arising as a special case. The latter case corresponds to the Israel-Wilson-Perj\'{e}s (IWP) metrics \cite{iwp}. The IWP solutions can be written in the form:
\bequ
\barr{l}
ds^2=-|H|^{-2}(dt+w_i dx^i)^2+|H|^2\delta_{ij}dx^i dx^j,
\\\\ F=\partial_i \mathcal{R}(H^{-1}) dt\wedge dx^i + \frac{1}{2}\left(|H|^2\delta^{kl}\epsilon_{ijk}\partial_l \mathcal{I}(H^{-1}) +2w_{i}\partial_{j}\mathcal{R}(H^{-1})\right)dx^i\wedge dx^j.
\label{iwpfiel}
\earr
\eequ
The latin indices denote spatial coordinates and take values $1,2,3$. We are using cartesian coordinates for the spatial metric. The complex function $H(x^i)$ is required to solve Laplace's equation in ${\Bbb E}^3$, and $|H|$ represents its modulus. It completely determines the solution since the rotation vector, $w^i$, is determined by
\bequ
\nabla \times {\bf w}=i(H\nabla H^*-H^*\nabla H),
\eequ
where the curl is taken for ${\Bbb E}^3$ and `*' denotes complex conjugation. The symbols $\mathcal{R}, \mathcal{I}$ denote, respectively, real and imaginary part of a complex quantity. 

There are two distinct families within the IWP solutions, according to how one complexifies $H(x^i)$ \cite{harhaw}. Complexifying the constants in the harmonic function leads to a multi-object generalisation of the charged Taub-NUT solution, whereas complexifying the coordinates in a specific way leads to the multi-object generalisation of the extreme (i.e. $Q^2=M^2$, where M,Q are the ADM mass and charge of the solution) Kerr-Newman (KN) spacetime. The one object limit of the latter - the usual KN solution - is obtained by choosing 
\bequ
H=1+\frac{M}{\sqrt{x^2+y^2+(z-ia)^2}},
\eequ
and contact with the standard form in Boyer-Linquist coordinates is made by changing from cartesian to oblate spheroidal coordinates, $(r,\theta,\phi)$, with transformations
\bequ
x+iy=\sqrt{(r-M)^2+a^2}\sin \theta e^{i\phi}, \ \ \ z=(r-M)\cos \theta,
\eequ
where the constant $a$ will be the angular momentum parameter.\footnote{Aside: Within the Kerr-Newman family, only the non-rotating case or the extreme $Q^2=M^2$ case can be expressed in the IWP form (\ref{iwpfiel}). To see this notice that a necessary condition is that after completing the square, $(dt+w)^2$, the remaining (i.e. transverse) metric must be conformally flat. Since the Weyl tensor is identically zero in three dimensions, it cannot be used to check conformal flatness. But there is a conformal tensor in three dimensions, firstly defined by Eisenhart \cite{eisen} but often called the York tensor \cite{york}, which is defined as
\bequ
Y_{abc}\equiv 2\nabla_{[c} R_{b]a}+\frac{1}{2}g_{a[c}\nabla_ {b]}R.
\eequ
The square parenthesis denote antisymmetrization with unit weight. For the Kerr-Newman spacetime the non-vanishing components for the York tensor applied to the transverse metric are proportional to $a^2(Q^2-M^2)$ as claimed.}

For the five dimensional case, not all supersymmetric solutions are known, but at least the equivalent to pp-waves and to the extreme Kerr-Newman are. The latter is known as the BMPV solution and given by the fields
\bequ
\barr{l}
\displaystyle{ds^{2}=-H^{-2}[dt+a_idx^i]^2+H\delta_{ij}dx^i dx^j,}
\\\\ \displaystyle{F=\frac{\sqrt{3}}{2}[\partial_i(H^{-1})dx^i\wedge dt+\partial_i(H^{-1} a_j)dx^i\wedge dx^j].}
\label{bmkn}
\earr
\eequ
The latin indices run now from 1 to 4. The solution is determined by \textit{two} independent quantities. The real function $H(x^i)$ is required to be harmonic on ${\Bbb E}^4$, and the ${\Bbb E}^4$ one form $a$ is required to have a Hodge self-dual field strength. In hyperspherical coordinates $(\rho,\theta, \phi_1, \phi_2)$, obtained from cartesian coordinates as 
\bequ
x^1+ix^2=\rho \cos\theta e^{i\phi_1}, \ \ \ \ x^3 +ix^4=-\rho \sin\theta e^{i\phi_2},
\label{hyper}
\eequ
the metric on ${\Bbb E}^4$ can be written as 
\bequ
ds_{{\Bbb E}^4}=d\rho^2+\rho^2(d\theta^2+\sin^2\theta d\phi_1^2+\cos^2\theta d\phi_2^2), 
\eequ
and the functions in (\ref{bmkn}) expressed as
\bequ
a=\frac{J}{2\rho^2}(\sin^2\theta d\phi_1 -\cos^2 \theta d\phi_2), \ \ \ \ \ \ H=1+\frac{\mu}{\rho},
\label{ah}
\eequ
where $J$, $\mu$ are constants.

Comparing the two solutions, the crucial difference is that the KN is determined by one complex function, whereas the BMPV is determined by one real function and a self dual form, independent of one another. For the latter spacetime, the gravitational potential, determined by the harmonic function, $H$, and the dragging effects, determined by the Sagnac connection, $a_i$, will be independent, while for the former spacetime they are interconnected. There are several consequences:
\begin{description}
\item[i)] The location of the null hypersurface defining the future event horizon, $\mathcal{H}^+$ (which, for these coordinates, is determined by the poles of the harmonic function), will depend on the angular momentum $J$ for the KN case, but not for the BMPV case. In particular we know that for the KN case only for $J=0$ will the harmonic function have real poles. In contrast the BMPV case will have a well defined horizon for non-vanishing $J$. The horizon becomes, however, ill-defined for large $J$, since the locus of $\mathcal{H}^+$ becomes a timelike hypersurface \cite{gibher};

\item[ii)] For the BMPV case, the norm of the timelike Killing vector field, does not depend on the angular momentum. Therefore, there cannot be an ergoregion and the angular velocity of the horizon must be zero. Since there is an overall angular momentum the spacetime cannot be rigidly rotating.

\item[iii)] This two cases are the only possibilities when trying to include rotation in a supersymmetric black hole spacetime: either we destabilise the horizon and create a naked singularity, or we keep a non-rotating horizon. A  rotating horizon is incompatible with supersymmetry since it will create an ergoregion \cite{GMTown}.
\end{description}

Comparing the two theories, two major differences are apparent: the dimension and the Chern-Simons term. The role of the Chern-Simons term was discussed in \cite{GMTown} where it was conjectured that the stability of the solution (and the regularity) of the horizon was associated with the presence of such term with the particular coefficient required by supersymmetry. But the particular spacetime dimension, $D=5$, was not sufficiently appreciated. A simple comparison might help. The similarity between the bosonic sector of $N=2, D=5$ and $N=8,D=11$ Supergravity has been long known, particularly the identical Chern-Simons terms \cite{cremmer}. The major difference is in the rank of the gauge field; a 1-form potential for the former versus a 3-form for the latter. They naturally couple to a black hole or a membrane. However, no regular, supersymmetric, rotating M-brane solution exists in the literature. In \cite{cve11}, the BPS limit of the non-extreme rotating M2-brane yields a spacetime with vanishing angular momentum (and an irregular horizon). Attempting to find such solution immediately reveals a major difference with the five dimensional case. In the derivation of \cite{GMTown} it was crucial to use the fact that the one form in the metric describing the rotation has a field strength with a self-duality property in the transverse space to the world volume (as will be review in section 4.2.1). As it stands, this property cannot be used in $D=11$, and singles out $D=5$ as a special dimension. 

We now give a simple example of how a self-duality condition can be useful in finding a solution with Killing spinors in a particular dimension. One can define a Brinkmann wave as a geometry that admits a covariantly constant null vector field: $N=\partial / \partial v$ obeying $\nabla_{\mu} N=0$. In general these waves can be described by a metric of the type
\bequ
ds^2=du(dv+f du +A_idx^i) + \hat{\gamma}_{ij}dx^i dx^j.
\label{brinkm}
\eequ
We are using light cone coordinates $u=-t+y$, $v=t+y$. Both the scalar function $f$, the spatial vector $A_i$ and the spatial metric $\hat{\gamma}_{ij}$ may admit $u$ and $x^i$ dependence. Pp-waves are the special case $A_i=0$ (or a pure gauge $A_i$, i.e., $A_i=\partial_i K$ for some function $K(x^i)$) and $\hat{\gamma}_{ij}=\delta_{ij}$. We will be interested in (\ref{brinkm}) with a flat spatial metric and no $u$ dependence. Then, demanding this geometry to be a solution to pure gravity (in arbitrary dimension) yields the constraints \cite{celes}
\bequ
\partial^k F_{ki}=0, \ \ \ \ \ \ \ \ \ \ \Delta f=-\frac{1}{8}F^{ij}F_{ij},
\label{riccif}
\eequ
where $F_{ij}=\partial_i A_j - \partial_j A_i$. If in addition one is looking for a supersymmetric configuration, the spacetime must possess a supercovariantly constant spinor, or equivalently for pure gravity, a covariantly constant spinor. Such metrics have a restricted holonomy and obey
\bequ
\partial_v \epsilon=0, \ \ \ \ \ \partial_u \epsilon +\frac{1}{16}F_{ij}\Gamma^{ij}\epsilon +\frac{1}{4}\partial_i f \Gamma^{u i}\epsilon=0, \ \ \ \ \ \partial_i \epsilon -\frac{1}{8}F_{ji}\Gamma^{uj}\epsilon=0.
\label{kilspi}
\eequ
For the pp-wave case, constant spinors with $\Gamma^{u}\epsilon=0$ (or, in $t,y$ coordinates, $(1+ \Gamma^{ty})\epsilon=0$) are solutions to (\ref{kilspi}) showing that these waves are one half supersymmetric. But for non trivial $A_i$ we have to deal also with the term $F_{ij}\Gamma^{ij}\epsilon$. A way to make this term vanish is to require the $x^i$ space to be four dimensional (${\Bbb E}^4$) and $F_{ij}$ to be the components of a Hodge self-dual two form in this space. We then find
\bequ
F_{ij}\Gamma^{ij}\epsilon=\frac{1}{2}F_{ij}\Gamma^{ij}(1+ \Gamma^{ty} \Gamma^7)\epsilon,
\eequ
where $\Gamma^7\equiv -\Gamma^{ty1234}$ is the chirality operator for the six dimensional spacetime. Thus, in $D=6$, (\ref{brinkm}) admits Killing spinors with one quarter of the degrees of freedom of a general constant spinor, due to the conditions
\bequ
\Gamma^7\epsilon =\epsilon, \ \ \ \ \ \ \ \ \ \ \Gamma^{u}\epsilon=0.
\eequ

The self-duality condition implies that the Maxwell equation in (\ref{riccif}) is identically obeyed. To solve the Poisson type equation we parametrise ${\Bbb E}^4$ with the hyperspherical coordinates (\ref{hyper}). Then, 
\bequ
A=\frac{J}{\rho^2}(\sin^2\theta d\phi_1 -\cos^2 \theta d\phi_2), \ \ \    
f=\frac{Q}{r^2}+\frac{J^2}{12r^6},
\label{abri}
\eequ
with $Q$ constant, is a solutions. Actually, the first term in $f$ could be any harmonic function on ${\Bbb E}^4$. We therefore find the following geometry to be a Brinkmann wave solution to $D=6$ pure gravity and admitting Killing spinors:
\bequ
ds^2=-dt^2+dy^2+\left(\frac{Q}{r^2}+\frac{J^2}{12r^6}\right)(dt-dy)^2+\frac{J}{r^2}(\sin^2\theta d\phi_1 -\cos^2\theta d\phi_2)(dy-dt)+ds_{{\Bbb E}^4}.
\label{brink6}
\eequ

Compactifying along the $y$ direction and performing Kaluza-Klein reduction, we find a 5D black hole type solution to Kaluza-Klein theory (in 5 dimensions). In the Einstein frame
\bequ
\barr{l}
\displaystyle{ds^2_{5,E}=-\Delta^{-\frac{2}{3}}\left[dt+\frac{J}{2\rho^2}(\sin^2\theta d\phi_1 -\cos^2 \theta d\phi_2)\right]^2+\Delta^{\frac{1}{3}}ds^2({\Bbb E}),}
\\\\ \displaystyle{A=-\left(1-\Delta^{-1}\right)dt+\frac{J}{2\rho^2\Delta}(\sin^2\theta d\phi_1 -\cos^2 \theta d\phi_2)},
\\\\ \displaystyle{e^{2\phi}=\Delta\equiv 1+\frac{Q}{\rho^2}+\frac{J^2}{12\rho^6}}.
\earr
\eequ
The solution has some unusual features, but we will postpone a detailed examination to somewhere else.

In section 3 we will need a 10D Brinkmann wave compatible with a system involving D1 and D5 branes. Such geometry will be similar to (\ref{brink6}) but without the $J^2/12r^6$ term. Some components of the Ricci tensor will then be non-zero, namely
\bequ
R_{tt}=R_{yy}=-R_{ty}=-R_{yt}=\frac{J^2}{r^8}.
\eequ
Therefore it is not a solution to pure gravity. It cannot be a solution to $D=6$, minimal supergravity either. In fact, the graviton multiplet is then $(g_{MN}, \Psi_{M}^{Weyl}, B_{MN}^{sd})$. The gravitino is a Weyl vector-spinor and the field strength $H_{MNP}^{sd}$ derived from the potential two form $B_{MN}^{sd}$ is required to be self-dual.\footnote{The self duality condition is crucial for the matching of on-shell degrees of freedom: $g_{MN}$ has $(D-2)(D-1)/2 -1=9$;  $\Psi_M^{Weyl}$ has $2^{\left[\frac{D}{2}\right]-1}(D-3)=12$; an usual two form potential $B_{MN}$ has $(D-2)(D-3)/2=6$, which can be halved by the self duality condition, obtaining the necessary matching for supersymmetry.} Hence, due to the Lorentzian $(-+++...)$ signature, $H_{MNP}^{sd}$ will have a vanishing energy momentum tensor, and any purely bosonic solution to simple $D=6$ supergravity must be Ricci flat.

In order to solve non trivially the Killing spinor conditions we required the space transverse to the Brinkmann wave propagation to be ${\Bbb E}^4$. However, we may add more flat directions, $z^{\alpha}$, without spoiling the supersymmetry, as long as the metric functions $A_i, f$ do not depend on $z^{\alpha}$. In this way we build the $D=10$ configuration:
\bequ
\barr{c}
\displaystyle{ds^2=du(dv+fdu+A_idx^i)+\delta_{ij}dx^i dx^j +\delta_{\alpha \beta}dz^{\alpha}dz^{\beta},}
\\\\ \displaystyle{B=\frac{A_i}{2}du \wedge dx^i,}
\label{brink10}
\earr
\eequ
with $\alpha=5...8$. Assuming self-duality of $F_{ij}$ in the aforementioned sense, (\ref{brink10}) is a supersymmetric configuration of $D=10$, type II supergravities.

Notice that the essential piece in this construction was to maintain an $SO(4)$ symmetry in the transverse space to the direction of propagation of the wave. This is an isometry of the space where the metric functions vary. The same isometry is present for a system of D1 branes inside D5-branes. Therefore we will be able to superimpose the systems and still have a supersymmetric solution. But the same is not true for the system of D2-D6-NS5 branes in type IIA that describe 4 dimensional black holes \cite{maldaphd}. From the higher dimensional viewpoint we can still see why we should expect different properties of 4 or 5 dimensional black holes.

For completeness we note that (\ref{brink10}) with $A$ given by (\ref{abri}) and  $f=C/r^2$ indeed solves the field equations for a graviton-axion configuration of $D=10$ type II supergravity, which read \cite{duffss}
\bequ
R_{\mu\nu}-\frac{1}{2}g_{\mu\nu}R=\frac{1}{4}\left(H_{\mu \sigma\tau}H_{\nu}^{ \ \ \sigma\tau}-\frac{1}{6}g_{\mu\nu}H_{\sigma\tau\rho}H^{\sigma\tau\rho}\right),\ \ 
D_{\mu} H^{\mu \sigma\tau}=0.
\eequ

Let us remark that the ten dimensional Brinkmann wave we have just described is not a particular case of the solutions derived in \cite{bko1}, due to the special dependence of the `rotation vector' on some of the coordinates of the transverse space rather than only one of the light cone directions.

\sect{Causality in Supergravity and String Theory}
In spite of fairly recent proposals for `infinitely large extra dimensions' \cite{ransun}, the most accepted way to match the observed number of dimensions with the theoretical requirements of string theory is still compactification. One assumes, therefore, a special topology for the universe, namely that it is a $K^6$ bundle over $M^4$, where $M^4$ is a four dimensional manifold and $K^6$ a compact six dimensional space. Compactification induces a lower dimensional spectrum consisting of an infinite tower of states. Most often only the massless ground states are of any relevance at low energies. These are related to the higher dimensional fields through the Kaluza-Klein procedure. When spacetime is a twisted $K^6$ bundle over $M^4$, the four dimensional metric obtained by the procedure will not be conformal to the 10 dimensional one, implying a modification in the causal structure: light cones for 10 and 4 dimensional gravity do not coincide anymore.\footnote{As an aside, let us remark on theories with variable speed of light. In fact, some recent attempts to deal with the cosmological problem of a (possibly) negative deacceleration parameter via gravity-scalar theories \cite{clamof} show a close technical similarity. In such theories, matter couples to gravity via a combination of the metric plus the derivatives of the scalar field. In our language it would be as if gravity was propagating in the whole ten dimensions, whereas matter would couple to a four dimensional Kaluza-Klein metric.}

In this section we analyse the distinctions between higher and lower dimensional viewpoints for the particular example of the BMPV black hole. To complete the picture we then review the string theory dual description of this spacetime \cite{BMPV}, and stress the relation between microscopic unitarity and macroscopic causality (which was implicit in \cite{BMPV}).

\subsection{Causality and dimensional reduction}

\subsubsection{The D1-D5-Brinkmann wave  system in D=10}
In 11 dimensions, a supersymmetric configuration was found describing the intersection of a five brane and a two brane on a string, with momentum along the string direction and with non-trivial angular momentum \cite{cve11}\footnote{The reader should notice that in formula (82) of this reference the expressions for $B^{(11)}_{t16}$ and $B^{11}_{t\phi_2 6}$ should read, respectively $-T$ and $J\cos^2\theta T/2r^2$.}. Using the following set of 11 dimensional coordinates we represent the configuration as:

\bigskip
\begin{center}
\begin{tabular}{|l|l|l|l|l|l|l|l|l|l|l|} \hline
t & $y_1$ & $y_2$ & $y_3$ & $y_4$ & $y_5$ & $y_6$ & r & $\theta$ & $\phi_1$ & $\phi_2$ \\ \hline
5 & 5 & 5 & 5 & 5 & 5 &&&&& \  \\
2 & 2 &&&&& 2 &&&& \\
  & w &&&&&&&&& \\ \hline
\end{tabular}
\end{center}
The last row refers to the spatial direction in which the wave carrying the momentum is propagating. Compactifying the $y^6$ direction to a circle and performing Kaluza-Klein reduction we obtain a solution of type II supergravity (IIA or IIB since no RR fields are excited). Transforming to the Einstein frame and S-dualizing we then get a solution to type IIB supergravity which can be interpreted as a $D1$-brane inside a $D5$ with a Brinkmann wave propagating along the string, as described in the last section:

\bequ
\barr{l}
\displaystyle{ds^2_E=f_5^{-\frac{1}{4}}f_1^{-\frac{3}{4}}\left[-dt^2+dy_1^2+f_K(dt-dy_1)^2 +\frac{J}{\rho^2}(\sin^2\theta d\phi_1-\cos^2\theta d\phi_2)(dy_1-dt)   \right]+}
\\ \ \ \ \ \ \ \ \displaystyle{+\left(\frac{f_1}{f_5}\right)^{\frac{1}{4}} ds^2({\Bbb E}_I^4)+f_5^{\frac{3}{4}}f_1^{\frac{1}{4}} ds^2({\Bbb E}_E^4)},
\\\\
\displaystyle{B_{RR}=-f_1^{-1}dt\wedge dy_1-P\cos^2\theta d\phi_1\wedge d\phi_2+\frac{J}{2\rho^2}f_1^{-1}\left(dy_1-dt\right)\wedge(\sin^2\theta d\phi_1-\cos^2\theta d\phi_2)},
\\\\
\displaystyle{e^{-2(\phi-\phi_\infty)}=\frac{f_5}{f_1}.}
\label{d15bri}
\earr
\eequ
Solutions with some similarities have been studied in \cite{papa}. 
Setting $J=0$ we recover the standard D1-D5-pp wave system \cite{calmal}. 
The Euclidean space ${\Bbb E}_I^4$ is parametrised by $y_2,y_3,y_4,y_5$, while the Euclidean space ${\Bbb E}_E^4$ is parametrised by $\rho, \theta, \phi_1, \phi_2$. The three functions $f_1$, $f_5$, $f_K$ are given by:
\bequ
f_5=1+\frac{P}{\rho^2}, \ \ \ \ \ \ \  \ \ f_1=1+\frac{Q}{\rho^2}, \ \ \ \ \ \ \ \ \ f_K=\frac{Q_{KK}}{\rho^2}.
\eequ
This solution is invariant under the transformations generated by four supercharges of the type IIB supersymmetry algebra. This corresponds to $1/8$ of the vacuum supersymmetry. Recall that the supersymmetry invariance of this theory is generated by two Weyl-Majorana spinors, $\epsilon_L$,$\epsilon_R$. The chirality conditions are $\Gamma^{11}\epsilon_{L,R}=\epsilon_{L,R}$. The Killing spinors obey three sets of conditions. The presence of the D-string and the D5-brane give rise to, respectively:
\bequ
\Gamma^{ty_1}\epsilon_L=\epsilon_{R}, \ \ \ \ \ \ \ \ \ \ \Gamma^{ty_1y_2y_3y_4y_5}\epsilon_L=\epsilon_{R},
\eequ
whereas the Brinkmann wave requires
\bequ
\Gamma^{ty_1}\epsilon_L=-\epsilon_{L},  \ \ \ \ \ \ \ \ \ \          \Gamma^{ty_1}\epsilon_R=-\epsilon_{R}.
\label{kilsbri}
\eequ
Of course, these are exactly the same conditions as in the $J=0$ case \cite{maldaphd}. The presence of angular momentum does not break any further supersymmetry, in the same way as it does not cost any more energy. However, as mentioned before, in solving the equation for the gravitino variation it becomes essential to use the fact that the one form representing the rotation, i.e. $A_i$ in $A_idx^idt$, is self-dual in the transverse space ${\Bbb E}_E^4$.

Compactifying further on $T^4\times S^1$, we obtain a five dimensional solution describing a rotating black hole with three different charges. Using U-duality arguments, one can quantise the charges $P,Q,Q_{KK}$ in terms of stringy quantities \cite{maldaphd}. Denoting the radius of the circle by $R$ and the volume of $T^4$ by $(2\pi)^4V$ the result is:
\bequ
P=Q_5 g \alpha', \ \ \ \ \ \ \ \ \ \ Q=\frac{Q_1\alpha'^3 g}{V}, \ \ \ \ \ \ \ \ \ \ Q_{KK}=\frac{g^2\alpha'^4 N_R}{R^2V}.
\eequ
The quantities $Q_1,Q_5$ and $N_R$ are integers and are counting, respectively, the number of D1-branes, D5-branes and units of right moving momentum, $g$ is the string coupling and $\alpha'$ the Regge slope. Unlike these quantities, the angular momentum parameter $J$ is still continuous. We will address its quantisation below.

\subsubsection{The BMPV black hole in D=5}
Let us describe the five dimensional configuration. The metric reads
\bequ
ds^2_E=-\left[f_1 f_5 (1+f_K)\right]^{-\frac{2}{3}}\left[dt+\frac{J}{2\rho^2}(\sin^2\theta d\phi_1-\cos^2\theta d\phi_2)\right]^2+\left[f_1 f_5 (1+f_K)\right]^{\frac{1}{3}}ds^2({\Bbb E}^4_E),
\label{bmpv3}
\eequ
while for the gauge fields we get
\bequ
\barr{c}
\displaystyle{A_i=-\frac{Q_i}{\rho^2+Q_i}dt+\frac{J}{2(Q_i+\rho^2)}(\sin^2\theta d\phi_1-\cos^2\theta d\phi_2)},
\\\\ \displaystyle{B=-P\cos^2\theta d\phi_1\wedge d\phi_2-\frac{J}{2(Q+\rho^2)}dt\wedge (\sin^2\theta d\phi_1-\cos^2\theta d\phi_2)},
\earr
\eequ
where $i=1,2$, $Q_1\equiv Q_{KK}$ and $Q_2\equiv Q$. Hence $A_1$ is the Kaluza-Klein gauge field arising from compactifying $y_1$ and $A_2$ is the winding gauge field arising from the same compactification. The moduli are
\bequ
e^{-2(\phi-\phi_\infty)}=\frac{f_5}{f_1}, \ \ \ \ \ \ \ \  e^{2\sigma_1}=\frac{1+f_K}{(f_5f_1^3)^{\frac{1}{4}}}, \ \ \ \ \ \ \ \ e^{2\sigma_s}=\left(\frac{f_1}{f_5}\right)^{\frac{1}{4}},
\eequ
where $s=2,3,4,5$. An equivalent solution was firstly obtained in \cite{cveyoum}. The ADM mass and the entropy of this black hole are 
\bequ
\barr{c}
\displaystyle{M_{ADM}=\frac{RV}{\alpha '^4 g^2}(Q+P+Q_{KK})=\frac{1}{g^2}\left(\frac{RgQ_1}{\alpha'}+\frac{RVgQ_5}{\alpha '^3}+\frac{g^2N_R}{R}\right)},
\\\\ \displaystyle{S_{Sugra}=\frac{\pi^2}{2G_5}\sqrt{PQQ_{KK}-\frac{J^2}{4}}}.
\label{massg}
\earr
\eequ
$G_5=\pi \alpha'^4 g^2/(4VR)$ is the five dimensional Newton's constant which relates to the ten dimensional one ($G_{10}=8\pi\alpha'^4 g^2$) by the moduli of the compact manifold. Notice that the ADM mass is unchanged from the static case as expected from supersymmetry. For the special case when all the three charges coincide, i.e. $Q_{KK}=P=Q\equiv \mu$, the configuration is equivalent to the one in \cite{BMPV}. The RR two form potential is then encoded in the remaining gauge fields through
\bequ
dB=\star d\bar{A}- \bar{A}\wedge d\bar{A},
\eequ
where $\bar{A}\equiv A_1=A_2$ and $\star$ denotes Hodge duality with respect to the metric (\ref{bmpv3}). All the scalars are constants. Therefore, in this special case, the solution is specified by the metric and the gauge field $\bar{A}$. In this form, the configuration can also be obtained as a solution to five dimensional $N=2$ supergravity \cite{GMTown}. Using a Schwarzchild type radial coordinate $r^2=\rho^2+\mu$, $g_E$ and $\bar{A}$ can be expressed as
\bequ
\barr{l}
\displaystyle{ds^{2}=-\left(\Delta_{10}\right)^2\left[dt+\frac{\mu\omega}{2(r^2-\mu)}(d\gamma+\cos{\beta}d\alpha)\right]^{2}+\frac{dr^{2}}{\left(\Delta_{10}\right)^2}+\frac{r^{2}}{4}[d\alpha^2+d\beta^2+d\gamma^2+2\cos{\beta}d\alpha d\gamma],}
\\\\ \displaystyle{A\equiv \frac{\sqrt{3}}{2}\bar{A}=-\frac{\sqrt{3}\mu}{2r^2}\left[dt-\frac{w}{2}\left(d\gamma +\cos\beta d\alpha \right)\right].}
\earr
\label{fields}
\eequ 
The angles $(\alpha,\beta,\gamma)$ are Euler angles on $SU(2)\simeq S^3$, we defined $w=-J/2\mu$ to make contact with \cite{gibher} and for future convenience we have introduced the notation
\bequ
\Delta_{ij}= 1-\left(\frac{\mu}{r^2}\right)^i\left(\frac{\omega}{r}\right)^{2j}.
\eequ
The configuration (\ref{fields}) is the BMPV black hole, which in isotropic coordinates becomes (\ref{bmkn}). We define the quantities $r_H$, $r_L$, $r_Q$ and $r_A$ as, respectively the zeros of $\Delta_{10}$, $\Delta_{21}$, $\Delta_{11}$, $\Delta_{01}$.  The first quantity defines the black hole horizon ($r=r_H$). The second quantity defines the surface $r=r_L$, which we call the Velocity of Light Surface (VLS). Inside the VLS there are closed timelike curves (CTC's).  We distinguish two cases. For $r_L>r_H$ there are naked CTC's. This is the over-rotating case. For $r_L<r_H$ the CTC's are hidden behind the horizon, which we refer to as the under-rotating case. Notice that 
\bequ
r_L\le r_H \Leftrightarrow  w^2\le \mu \Leftrightarrow \frac{J^2}{4}\le Q_{KK}QP,
\label{caubou}
\eequ
which is therefore the causality bound. We remark that the interpretation of $r=r_H$ as the black hole horizon only makes sense in the under-rotating case, since the horizon should be a null hypersurface which is no longer true in the over-rotating case. The interpretation of $r_Q$ and $r_A$ will be given in section 4.1.

At this point we should comment on some conventions. The standard treatment of $N=2$, $D=5$ Supergravity leads to a Bogomol'nyi bound of $M\ge \sqrt{3}|Q|/2$ \cite{traschen, GMTown}. This leads to a factor of $\sqrt{3}/2$ in the gauge potential, so that the field $A$ introduced in (\ref{fields}) is consistent with this treatment. This factor is then essential for supersymmetry. The potential $\bar{A}$ coming from dimensional reduction is consistent with a bound without the $\sqrt{3} /2$ factor. Therefore, one must perform a rescaling of the field to make contact with the five dimensional theory. Although using $A$ leads to a more awkward set of conventions, we will use it when dealing with the five dimensional viewpoint.

\subsubsection{Causality}
The geometry (\ref{fields}) contains strong causality violations. Even a freely falling particle can move backwards in time (as seen from the observer at infinity) \cite{gibher}. What happens in $D=10$?

The geometry (\ref{d15bri}) does not have any `obvious' closed timelike curves, in the sense that there is no periodic direction whose metric coefficient changes sign in some spacetime region. However if we choose to compactify the $y_1$ direction, we create CTC's. The point is that there are linear combinations like $t^{\mu}\partial_{\mu}=2B/r(\partial/ \partial \gamma) +A(\partial /\partial y_1)$ that become timelike (we have chosen to parametrise the ten dimensional solution with Euler angles for these remarks). But only when the $y_1$ direction is made periodic, may the curves with such tangents become closed.
Rewriting the $10D$ configuration with all the charges equal and changing to Schwarzchild coordinates as before, the condition for such curves to become null is:
\bequ
\left|\frac{B}{A}\right|=\left(\frac{r_L}{r}\right)^3\pm\sqrt{-\Delta_{21}}.
\eequ
Only for $r<r_L$ two distinct roots arise. Values of $|B/A|$ in between them correspond to timelike curves. This happens only inside the VLS, which can therefore be seen from the higher dimensional perspective as well. The condition for the curves to be closed is 
\bequ
\frac{B}{A}=\frac{r}{2R}q,
\eequ
where $q\in {\Bbb Q}$.

The universal covering space of the manifold with geometry (\ref{d15bri}) is therefore causally well behaved, whereas its non-simply connected compactification has CTC's. Notice that this is quite different from creating causal anomalies in flat space by identifying the time coordinate or from the $AdS$ causal problems. In both these cases, it is a timelike direction that becomes compact, whereas in our case CTC's become possible through the compactification of a  spacelike direction. The similarity is, of course, they both can be resolved by going to the universal covering space.

The Kaluza-Klein reduction to five dimensions gives rise to the identifications
\bequ
g_{MN}^{(10,E)} = \left( \barr{c} g_{\mu \nu}^{(5,KK)}+e^{2\sigma_1}A_{\mu}A_{\nu} \ \ \   \ \ \  e^{2\sigma_1}A_{\mu}  \ \ \   \ \ \  0 \\ \ \ \  \ \ \ \ \ \ \ e^{2\sigma_1}A_{\nu} \ \ \ \ \ \  \ \ \ \    \ \ \ \ \    e^{2\sigma_1}    \ \ \ \ \   \ \ \ 0 \\ \ \ \ \ \ \ \ \ \ \ \ \ \ \ \ 0  \ \ \ \ \ \ \ \ \ \ \ \ \ \ \ \  \ \ \ \ \ \ \ 0 \ \ \ \ \ \ \  \ \ \ e^{2\sigma_s} \earr \right).
\label{kkframe}
\eequ 
The superscripts on the metric refer to the Einstein or `Kaluza-Klein' frame\footnote{By Kaluza-Klein frame we mean the frame obtained from a Kaluza-Klein compactification starting from the higher dimensional Einstein frame and without rescaling the lower dimensional metric by using the moduli, as in using ansatz (\ref{kkframe})}. The first row (and column) refers to the directions $t$ and ${\Bbb E}_E^4$, the second to the direction $y_1$ and the last to the directions in ${\Bbb E}_I^4$. The five dimensional Einstein frame is then obtained as 
\bequ
g^{(5,E)}=e^{\frac{2}{3}\sum_{i=1}^5 \sigma_i}g^{(5,KK)}.
\eequ
The procedure eliminates the $y_1$ direction and effectively projects down the CTC's into the $\gamma$ direction. So, the metric $g_{\mu \nu}^{5,KK}$ (or $g_{\mu \nu}^{5,E}$) has unavoidable closed timelike curves, whereas the combination 

\bequ
g_{\mu \nu}^{(5,KK)}+e^{2\sigma_1}A_{\mu}A_{\nu} 
\label{regul}
\eequ
does not. The latter describes the local geometry of a manifold were any existing CTC's will not be homotopic to a point.

\begin{figure}
\begin{picture}(0,0)(0,0)
\put(67,18){$\rho$}
\put(385,55){$\rho$}
\put(210,5){$\partial /\partial \gamma$}
\put(40,50){$t^{\mu}\partial_{\mu}$}
\put(95,35){$\partial /\partial y_1$}
\put(189,36){$\partial /\partial y_1$}
\put(260,40){$t^{\mu}\partial_{\mu}$}
\put(33,5){$\partial /\partial \gamma$}
\put(362,19){$\partial /\partial \gamma$}
\end{picture}   
\centering\epsfig{file=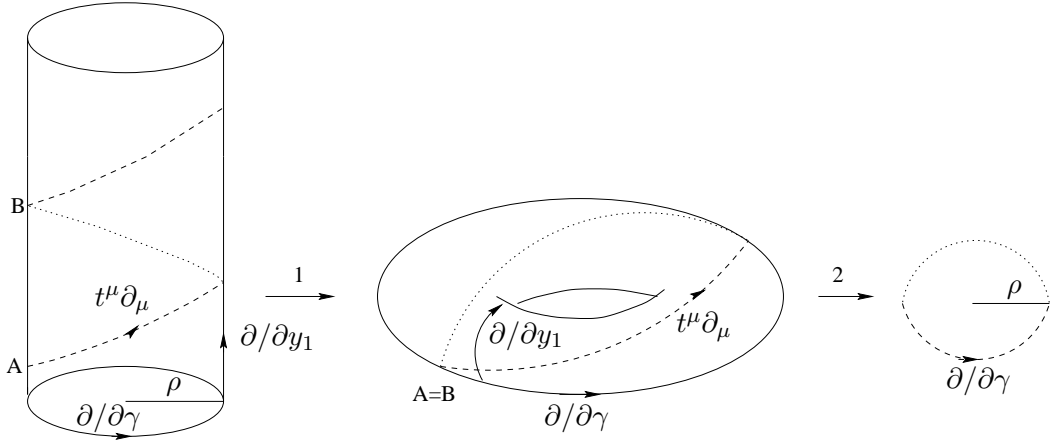,width=14cm}   
\caption{Creation of CTC's for the BMPV spacetime from the $D=10$  causal configuration.}
\label{ctccre}
\end{figure}

In Figure \ref{ctccre} we illustrate the above procedure. The picture on the left describes the $D=10$ universal covering manifold, i.e., $y_1\simeq {\Bbb R}$, and we illustrate the curve with tangent $t^{\mu}\partial_{\mu}$. Step $1$ is the compactification of $y_1$, i.e., $y_1\simeq S^1$, so that the curve with tangent $t^{\mu}\partial_{\mu}$ becomes closed. Step $2$ is the Kaluza-Klein reduction, upon which the curve is projected onto the $\gamma$ direction, corresponding to the causal anomalies seen in the BMPV black hole. We should remark that the pictures are misleading in two senses. Firstly, the manifold parametrised by $y_1$ and ${\Bbb E}^4_E$ is a non trivial ${\Bbb R}$ (or $S^1$) bundle over ${\Bbb E}^4_E$, which is precisely what allows the curve described by $t^{\mu}\partial_{\mu}$ to become timelike. Secondly, the $\gamma$ direction does not go around a non-trivial cycle of the manifold.

Can we associate other acausal spacetimes with better behaved higher dimensional configurations in a similar fashion? This would rely on performing a transformation of the type (\ref{regul}). Thus, one cannot apply the method to any acausal spacetime configuration, since not all examples have a gauge field. The G\"{o}del manifold is the simplest example \cite{godel}. But it turns out that this procedure does not work for the general five dimensional family of black holes \cite{cveyoum}. Moreover it is not something associated to supersymmetry either, since the four dimensional Kerr-Newman spacetime with $Q^2=M^2$  still exhibits the traditional closed timelike curves in the negative $r$ region after applying an analogous transformation to (\ref{regul}). Therefore, we have to conclude that this spacetime has a peculiar property, which, using Carter's terminology \cite{carter}, one can put as: the non-trivial CTC's of the five dimensional spacetime arise as trivial CTC's from the ten dimensional viewpoint.

\subsection{Causality and Unitarity}
Consider the ten dimensional supergravity D1-D5-brane solution, i.e., (\ref{d15bri}) with $Q_{KK}=J=0$, and $y_1$, ${\Bbb E}_I^4$ compact.  The D-brane tensions depend on the string coupling constant $g$ as $1/g$ (as can be seen from (\ref{massg})) and the ten dimensional Newton's  as $G_{10}\sim g^2$. Hence, the gravitational potential for D-branes goes as $V\sim g$. Decreasing the string coupling we turn off gravity and find a flat space configuration: a gas of open strings with boundary conditions determined by the presence of the D-branes.

The theory describing these open strings and effectively describing the D-branes is a 1+1 dimensional SCFT (Super Conformal Field Theory), which can be thought of as living on the worldvolume of the D1-branes inside the D5-branes \cite{vafa}. Within the AdS/CFT correspondence this is because the near horizon geometry of the $D=10$ configuration is $AdS_3\times S^3\times T^4$, and the 1+1 SCFT lives on the conformal boundary of the AdS piece. Since we compactified the supergravity configuration to 5D, our SCFT will be on the cylinder ${\Bbb R}\times S^1$, were ${\Bbb R}$ is the time direction. The amount of supersymmetry of this SCFT must be the same as in the classical configuration, i.e., $1/4$ of the vacuum supersymmetry. Thus, we have an $N=4$ SCFT on  ${\Bbb R}\times S^1$ describing the compactified supergravity D-brane configuration.

Supersymmetric states belong to short multiplets. These should be stable at any value of the string coupling. Therefore we expect our continuous variation of $g$ to leave the degeneracy of states unchanged, i.e., to be adiabatic. The SCFT has a central charge determined by the massless degrees of freedom of the open strings, $\tilde{c}=N_B+N_F/2$.\footnote{In our setup the states describing the entropy are right moving. That is why the central charge has a tilde. But this is the usual Virasoro central charge, related to the complex dimension $d_c$ of the target space for the sigma model by $\tilde{c}=3d_c$.} These are fairly easy to count: the massless excitations have the same number of bosonic ($N_B$) and fermionic ($N_F$) physical degrees of freedom, $4Q_1Q_5$, effectively carried by the (1,5) and (5,1) strings in the flat space D-brane configuration \cite{calmal}. Hence $\tilde{c}=6Q_1Q_5$.

Now we want to turn on momentum ($Q_{KK}$) and angular momentum ($J$) in the supergravity side and see the correspondence in the SCFT side.

The supergravity angular momentum is classified by the $SO(4)$ rotation group which is an isometry of the geometry (\ref{d15bri}) or (\ref{bmpv3}) acting on the space transverse to the branes. We have to identify this symmetry on the SCFT side. It is known that an $SO(4)$ symmetry arises in the $N=4$ superconformal algebra, corresponding to endomorphisms in the graded algebra that rotate the fermionic generators $G_m^i$ (i=1..4) amongst themselves, i.e. R-symmetry. Such symmetry is gauged in the sense that linear combinations of the $G_m^i$ (for negative m) create states that carry $(F_L,F_R)$ charges of the $U(1)_L\times U(1)_R$ Cartan subalgebra of $SO(4)$. These two quantities are therefore to be identified with the two linearly independent spacetime angular momentum parameters
\bequ
(J_L,J_R)\equiv (J^{21}-J^{43}, J^{21}+J^{43})=(0, \frac{\pi}{4G_5}J)=(0, \frac{VR}{\alpha'^4 g^2}J).
\eequ
The $J^{ik}$ is the angular momentum tensor that can be read off from the geometry (\ref{bmpv3}) in the usual way. In this way, the $J_L$, $J_R$ become quantised in terms of $(F_L,F_R)$. Using the above results for the quantisation of $P,Q,Q_{KK}$, we may now express the entropy of the classical solution in terms of quantised charges:
\bequ
S_{Sugra}=\frac{\pi^2}{2G_5}\sqrt{PQQ_{KK}-\frac{J^2}{4}}=2\pi\sqrt{Q_1Q_5N_R-\frac{F_R^2}{4}}.
\eequ
We have not identified $N_R$ in the SCFT side yet, i.e., established the correspondence between SUGRA momentum and a quantity in the SCFT. In supergravity the integer $N_R$ arises due to the compactification of the $y_1$ direction into $S^1$. It is therefore a $U(1)$ charge. The conformal algebra of the cylinder - the Virasoro algebra - has certainly a $U(1)$ subalgebra. States carry a $U(1)$ charge given by the eigenvalue of the zero mode in the theory, i.e., the Virasoro generator $\tilde{L}_0$. Hence, $N_R$ must correspond in the SCFT to the right moving level $n_R$ of the states.

What is the degeneracy of states of the form 
\bequ
|n_L,F_L;n_R,F_R>=|0,0;N_R,F_R> 
\label{states}
\eequ
in an $N=4$ SCFT of central charge $\tilde{c}=6Q_1Q_5$? For a 2D CFT on a cylinder, the degeneracy of states at highest available level $M>>1$ can be computed by using the modular invariance of $T^2$ in the Euclidean section \cite{cardy}, yielding, up to power corrections
\bequ
d(M,\tilde{c})\sim e^{2\pi\sqrt{\frac{M\tilde{c}}{6}}}.
\eequ
The highest level available for the states (\ref{states}) is not $N_R$, after we have fixed the other quantum numbers, namely $F_R$. In a CFT, states associated with a primary operator give a contribution to the total $\tilde{L}_0$ eigenvalue equal to their conformal weight. In particular, an operator crating states with charge $F_R$ has a conformal weight not smaller than $3F^2_R/(2\tilde{c})$ \textit{if we assume} that the CFT is \textit{unitary}, which demands all conformal weights to be non-negative \cite{pol}. The total $\tilde{L}_0$ eigenvalue, $N_R$, is therefore the sum of $3F^2_R/(2\tilde{c})$ plus the conformal weight of other operators, $N_R^{remain}$, which is non-negative by unitarity:
\bequ
N_R=\frac{3F^2_R}{2\tilde{c}}+N_R^{remain} \Rightarrow Q_1 Q_5 N_R\ge \frac{F^2_R}{4}.
\label{unibou}
\eequ
This is the unitarity bound, which coincides with the causality bound (\ref{caubou}). Moreover, the highest available level is now $M=N_R^{remain}$. This is the total level available to find as different combinations of operators. It therefore determines the degeneracy of states. It follows that 
\bequ
S_{SCFT} = \ln{d(N_R^{remain}, \tilde{c})} = S_{Sugra}
\eequ

The above bound can also be seen in the explicit constructions of unitary representations of the SCFT. First notice that the states we are interested in are in the left-moving ground state. So, we need to look only at the representation theory of an $N=2$ SCFT, which was worked out in \cite{kent}. The generators are $L_m$, the Virasoro generators, $T_m$, the modes of a $U(1)$ current, and $G_m^i$, $i=1,2$, the modes of the supercurrent. The $U(1)$ symmetry is the only surviving piece of the $SO(4)$ symmetry of the $N=4$ algebra, which we want to think of as $U(1)_R$. There are 3 possible N=2 algebras, according to the moddings one can choose for the generators, via the boundary conditions. We are interested in the P algebra, where the fermions have periodic boundary conditions and therefore generalises the usual Ramond sector of the open string theory. The reason is that the spacetime angular momentum is carried by the fermionic modes of the (1,5) and (5,1) strings \cite{maldaphd}. 

There are 3 possible classes of unitary representations for the P algebra. The ones with a two dimensional moduli space obey \cite{kent}\footnote{Note: $\tilde{c}$ in \cite{kent} is the complex dimension of the target space, hence 3 times our usual Virasoro central charge.}
\bequ
2\left(\frac{\tilde{c}}{3}-1\right)\left(h-\frac{\tilde{c}}{24}\right)-q^2+\frac{1}{4}\left(\frac{\tilde{c}}{3}+1\right)^2\ge0,
\eequ
which, for $\tilde{c}=6Q_1Q_5>>1$, level $h=N_R$ and charge $q=F_R$ gives (\ref{unibou}).

The upshot is that the charge (in CFT language) or angular momentum (in spacetime language) we create must be bounded by the energy we create; otherwise we violate unitarity or causality.

\sect{Other Properties of the BMPV spacetime}

\subsection{The `repulson' behaviour}
The first property we would like to address is related with the over-rotating case. Let us start by computing the charged geodesics. We follow the Hamilton-Jacobi method with the usual minimally coupled Hamiltonian 
\bequ
H=\frac{g^{\mu\nu}}{2}\left(p_{\mu}+qA_{\mu}\right)\left(p_{\nu}+qA_{\nu}\right)
\eequ
where $g_{\mu \nu}$ and $A_{\mu}$ are given by (\ref{fields}). The ansatz for the action function is
\bequ
S=-Et +H(\alpha, \beta, \gamma)+W(r), \ \ \ \ H(\alpha, \beta, \gamma)=j_L \alpha +j_R\gamma +\chi (\beta).
\eequ
This generalises the construction in \cite{gibher}. The Killing tensor found there is reducible. Since each of the Killing vector fields into which it decomposes is a  symmetry  of the Maxwell field (in the sense that $\pounds_{L_3}A=\pounds_{R_i}A=0$, where $\pounds$ denotes the Lie derivative and $L_i (R_i)$ the left (right) invariant vector fields on $SU(2)$), we conclude that the Killing tensor still commutes with the minimally coupled Hamiltonian. Therefore the quantity
\bequ
j^2\equiv \left(L_1 H\right)^2+\left(L_2 H\right)^2+\left(L_3 H\right)^2,
\eequ
is still a constant of motion, as in the case of purely gravitational interactions. The following set of equations of motion is then obtained (for the BMPV black hole, the mass to charge ratio is $M/Q=\sqrt{3}/2$, but we keep $Q$ and $M$ in order to make the interpretation of the several terms clear):
\bequ
\barr{c}
\displaystyle{\left(\frac{dr}{d\lambda}\right)^2=E^2\Delta_{21}-\left(m^2+\frac{4j^2}{r^2}\right)\left(\Delta_{10}\right)^2-\frac{qQE}{r^2}\Delta_{11}+\left(\frac{qQ}{2r^2}\right)^2 \Delta_{01} +\frac{2\omega j_R }{r^4}\left(qQ-\frac{4}{3}ME\right)\Delta_{10}},
\\\\ \displaystyle{\frac{dt}{d\lambda}=\frac{1}{\left(\Delta_{10}\right)^2}\left[E\Delta_{21}-\frac{qQ}{2r^2}\Delta_{11}-\frac{4\omega M j_R}{3r^4}\Delta_{10}\right]},
\\\\ \displaystyle{\frac{d\gamma}{d\lambda}=\frac{4}{r^2}\left[\frac{j_R-j_L\cos\beta}{\sin^2\beta}+\frac{\omega}{4r^2\Delta_{10}}\left(\frac{4}{3}ME -qQ\right)\right].}
\earr
\eequ
The $\beta$ and $\alpha$ equations of motion are the same as in the purely gravitational case \cite{gibher}. The quantity $m^2$ is the mass squared of the test particle, which arises in the usual fashion as the integral of motion associated with the metric Killing tensor. Notice that the RHS of the $r$ equation is even under CPT, whereas the RHS of the $t$ and $\gamma$ are odd, as expected from the form of the LHS. Also notice that the affine parameter $\lambda$ is not proper time $\tau$ for the massive case, but $\tau=\lambda m$, and $m$ can be made to vanish in the equations by defining energy, charge and angular momentum per unit mass. 

We now discuss some features of these equations. There are four interesting surfaces: the horizon (at $r=r_H$), the VLS (at $r=r_L$) and the timelike surfaces $r_Q$, $r_A$, defined as the zeros of $\Delta_{11}$ and $\Delta_{01}$ respectively. The relative location of these surfaces is $r_A<r_Q<r_L<r_H$ ($r_A>r_Q>r_L>r_H$) for the under-rotating (over-rotating) case. The VLS bounds the region where light cones allow causal travelling into the past of the observer at infinity. The energy term in the $t$ equation changes sign corresponding to the possibility of geodesic time travelling, or equivalently, to a change of character from particle to antiparticle. The energy term in the radial equation also changes sign and becomes repulsive. The surface $r=r_Q$ couples to the Coulomb term both in radial and time equations. Both of them change signs when crossing this surface, i.e., the Coulomb interaction changes from attractive to repulsive or vice-versa. We call this surface `Coulomb Conjugation Surface (CCS)'. The $r=r_A$ surface corresponds to a change in sign for the general relativistic correction to the electromagnetic interaction (which is asymptotically subleading). Effectively $q^2$ changes sign there. 

On the surface $r=r_H$, the right hand side of the $r$ equation of motion becomes
\bequ
\left(E-\frac{3qQ}{4M}\right)^2\left(1-\left(\frac{r_L}{r_H}\right)^6\right).
\label{rhori}
\eequ
This is always negative in the over-rotating case. We conclude, therefore, that the repulson like behaviour found in \cite{gibher} for the gravitational interactions is still true in general when charge interactions are taken into account: for the over-rotating solution there are no freely falling orbits, with or without charge, entering the $r=r_H$ surface. The case singled out by (\ref{rhori}) corresponds to an energy to charge ratio for the  test object of $\sqrt{3} / 2$. With our conventions, within a supersymmetric theory $E\ge m \ge \sqrt{3}/2|q|$. Thus a particle for which (\ref{rhori}) vanishes everywhere should have $E=m$, i.e. be at rest, so the mass to charge ratio is also $\sqrt{3}/2$- a BPS particle. The behaviour is then better seen by rewriting the $r$ equation of motion:
\bequ
\barr{l}
\displaystyle{\left(\frac{dr}{d \lambda}\right)^2 = \left(qQ-\frac{4}{3}ME\right)\left[\frac{2\omega j_R}{r^4}\Delta_{10}-\frac{\omega^2}{4r^6}\left(qQ-\frac{4}{3}ME\right)\right]+}
\\\\ \ \ \ \ \ \ \ \ \ \ \  \displaystyle{+\left(E^2-m^2\right)-\frac{1}{r^2}\left(qQE-\frac{4}{3}Mm^2\right)+\frac{1}{4r^4}\left((qQ)^2-\frac{16}{9}(mM)^2\right)-\frac{4j^2}{r^2}\left(\Delta_{10}\right)^2.}
\label{rbps}
\earr
\eequ
Since the BPS test particle must have for initial conditions the lowest possible quantum numbers, $|E|=m, j=0$, it will stay at rest in the spatial coordinates, i.e. a no-force configuration.

The repulson result is readily confirmed at the semi-classical level for charged scalar waves minimally coupled to the background electromagnetic field. We also introduce a non-minimal coupling to the geometry. We are still able to separate variables in the latter model, which follows from the Ricci scalar being a function of $r$ only:
\bequ
R=\frac{2}{r^8}(2\mu^2\omega^2-r^2\mu^2).
\eequ
The wave equation then takes the form

\bequ
g^{\mu \nu}(D_{\mu}+iqA_{\mu})(D_{\nu}+iqA_{\nu})\Phi=(m^2+\lambda R)\Phi.
\label{scar}
\eequ
The non-minimal coupling might be thought of as a renormalisation of the mass. Following \cite{gibher} we use the ansatz $\Phi(x^{\mu})=e^{-itE}D^{j}_{j_L,j_R}F(r)$ and change variables as 
\bequ
x=\frac{\mu}{r^2-\mu},
\eequ
in which case (\ref{scar}) can be written as
\bequ
\frac{d^2 F}{dx}=\left[A+\frac{B}{x}+\frac{C}{x^2}+\frac{D}{x^3}-\frac{\lambda}{x^2}\left(\frac{\omega^2x^3}{\mu(1+x)^3}-\frac{x^2}{(1+x)^2}\right)\right]F.
\label{scax}
\eequ
A,B,C,D are given by
\bequ
\barr{c}
\displaystyle{A=\frac{1}{4\mu}\left(\left(\frac{r_L}{r_H}\right)^6-1\right)\left(E\mu-\frac{qQ}{2}\right)^2, \ \ \ \ \  C=D+j(j+1)+\frac{E}{2}\left(\frac{qQ}{2}-\mu E\right),} \\
\displaystyle{B=\frac{1}{4\mu}\left(\left(\frac{qQ}{2}\right)^2-\mu^2E^2\right)-\frac{1}{2\mu}\left(\mu E -\frac{Qq}{2}\right)^2-\frac{\omega j_R}{2\mu}(qQ-2\mu E), \ \ \ \ \ D=\frac{\mu}{4}(m^2-E^2).}
\earr
\eequ
Near the horizon the scalar equation (\ref{scax}) is approximated by keeping only the $A$ term on the right hand side. Oscillating solutions arise only if $r_L<r_H$, i.e. in the under-rotating case. Therefore the absorption cross section for scalar charged waves is zero in the over-rotating case,  confirming the `repulson' behaviour.

Now we ask if an accelerated observer can enter the `horizon' in the over-rotating case. Obviously, there are timelike curves that can achieve that. The simplest example is a radial orbit with tangent vector
\bequ
t^{\mu}\partial_{\mu}=\frac{\sqrt{2}}{\Delta_{10}}\partial_t-\Delta_{10} \partial_r,
\label{radac}
\eequ
which has a unit norm. But it is well known in general relativity that a generic timelike trajectory might not be `realistic'. The simplest example is given by the Reissner-N\"ordstrom spacetime. The repulsive character of the timelike singularity precludes any timelike trajectory with bounded proper acceleration to reach the physical singularity at $r=0$ (in Schwarzchild type coordinates) \cite{accgr}. The non real character of such curves is manifest in the fact one would need an infinite acceleration (and so an infinite payload for the rockets) to perform such a deed (even assuming one could survive the tidal forces). In the  Reissner-N\"ordstrom case, the divergence in the proper acceleration is a consequence of the existence of an s.p. singularity. In our case, however, the most obvious curvature invariants ($R$,$R_{\mu \nu}R^{\mu \nu}$, $R^{\mu\nu\sigma\tau}R_{\mu\nu\sigma\tau}$) show no such singular behaviour on or outside the $r=r_H$ surface. Furthermore, we could not find any pp-singularity.\footnote{Black holes with this kind of singular behaviour outside the horizon -`naked black holes'- were studied in \cite{horw}.} 

What happens to an accelerated observer? Along the trajectory described by (\ref{radac}), the proper acceleration gives
\bequ
a^{\mu}a_{\mu}=t^{\nu}D_{\nu}t^{\mu}t^{\tau}D_{\tau}t_{\mu}=\frac{8\mu^2}{r^6}\left( 1+\frac{w^2}{r^2}\right).
\eequ
There is no divergence here. Therefore it seems one would be able to travel into the $r<r_H$ region if (and only if) one would have a rocket and a (very) robust spaceship. This is clearly an odd behaviour.

Let us conclude this section by commenting on some recent work on repulsons. As mentioned above, the RN curvature singularity is repulsive. In particular, the extreme RN spacetime with negative mass has a repulsive naked singularity. This is the type of repulsons dealt with in \cite{peetpol}. In order to make contact with the AdS/CFT conjecture, the repulsons therein are brane configurations with near horizon geometries containing an AdS piece. One example is (\ref{d15bri}) with $J=Q_{KK}=0$ and $Q<0$. Others may be obtained by T-duality. These repulsons are therefore quite different from our case, since the repulsive hypersurface is non-singular for the over-rotating BMPV spacetime.

\subsection{The gyromagnetic ratio for the BMPV black hole}
We also want to address the gyromagnetic ratio of these black holes. As seen before, from the asymptotic behaviour of the metric (\ref{fields}), we can read off how the two parameters $\mu$ and $\omega$ are related to the ADM mass and the two angular momenta parameters of the $SO(4)$ rotation group:
\bequ
M_{ADM}=\frac{3\pi \mu}{4 G_5}, \ \ \ \ \ \ (J_L,J_R)=(0,-\frac{\pi \mu \omega}{2G_5}).
\eequ
To get the latter expression we can use the relation between Euler angles and the cartesian coordinates on ${\Bbb E}^4$, $X^i$. Such relation is obtained  by solving the embedding constraints for the $S^3$ embedding in ${\Bbb E}^4$. We can then write
\bequ
d\gamma+\cos\beta d\alpha=\frac{2(L^3)_{ki}X^k dX^i}{\rho^2},
\label{form}
\eequ
where $L^3$ is a Hodge self dual two form on ${\Bbb E}^4$, and $\rho=\sqrt{X^i X_i}$. The spacetime angular momentum can now be expressed as
\bequ
J^{ki}=\frac{\pi\mu\omega}{4G_5} (L^3)^{ki},
\eequ
and our definitions of left and right angular momenta are $J^{21}\pm J^{43}$, with the `+' sign for $J_R$.

The charge associated with the Maxwell field is given by\footnote{An alternative quite frequent charge normalization is $Q=(2 A_{D-2} G_D)^{-1}\oint_{S^{D-2}} F^{\mu \nu}dS_{\mu \nu}$. Then, the charge, magnetic dipole moment and coefficient in Bogomol'nyi bound will differ from ours by a factor of $\pi / 2$. In this section we set $G_5=\pi / 2$. So our formulae are the same as the ones we would obtain with the alternative charge definition and setting $G_5=1$.} 
\bequ
Q=\frac{1}{8\pi G_5} \oint_{S^3}F^{\mu\nu}dS_{\mu \nu}=-\frac{\sqrt{3}\mu\pi}{2G_5},
\label{charge}
\eequ
and so the mass to charge ratio is $\sqrt{3}/2$, i.e. it saturates the Bogomol'nyi bound for the $D=5$, $N=2$ supergravity theory \cite{traschen}.  We should remark that the normalization in (\ref{charge}) and for the ADM mass are consistent with Newtonian and Coulomb force laws of the form
\bequ
\|{\bf F_N}\|=\frac{8G_5}{3\pi}\frac{M_1 M_2}{r^3}, \ \ \ \  \|{\bf F_C}\|=\frac{2G_5}{\pi}\frac{Q_1 Q_2}{r^3},
\eequ
which means that a gravitational-electrostatic force balance is achieved when $4 M_1 M_2 =3 Q_1 Q_2$. An object experimenting a no-force condition on a BPS background is then expected to possess the same mass to charge ratio of $\sqrt{3} / 2$, as seen in the last section.

Rotation endows a charged black hole with a magnetic dipole moment, which can be read off from the spatial components of the vector potential. Using (\ref{form}) we get
\bequ
\mu^{ij}=-\frac{\sqrt{3}\pi\mu\omega}{4G_5}(L^3)^{ij},
\eequ
and from the usual relation we can read off the gyromagnetic ratio:
\bequ
\mu^{ij}=g\frac{Q}{2M}J^{ij} \ \ \Rightarrow \ \ \  g=3.
\eequ
Notice we would obtain the same value if we had used $\bar{A}$ in (\ref{fields}) instead, since an overall factor gives the same contribution to charge and magnetic moment.

\subsubsection{Black Hole superpartners}
The field configuration (\ref{fields}) can be expressed very simply using an isotropic type radial coordinate, $\rho$ defined as $\rho^2+\mu=r^2$. Then we have (\ref{bmkn}). The harmonic form $H$ and the one-form $a$ are given by (\ref{ah}) for a spacetime with a connected even horizon. The multi black hole configuration is obtained by using a harmonic function with $N$ poles:
\bequ
H(x^i)=1+\sum_{\alpha=1}^{N} \frac{\mu_{\alpha}}{|x^i-x^i_{\alpha}|^2}, \ \ \ \ \ a=\frac{J}{4\mu}(L^3)^k_{\ i}\partial_k H dx^i,
\eequ
where $x^i_{\alpha}$ are constants. A quite natural set of frames is
\bequ
e^0=H^{-1}(dt+a), \ \ \ \ \  e^i=H^{\frac{1}{2}}dx^i,
\eequ
for which the field strength $F$ takes the form $F=\sqrt{3} d (e^0) /2$. Using the Cartan structure equation for the spin connection and expressing $F$ and the exterior derivative in terms of the frames we get \cite{GMTown}:
\bequ
\barr{c}
\displaystyle{\omega^0_{\ \ i}=-\frac{\partial_i H}{H^{\frac{3}{2}}}e^0 +\frac{f_{ij}}{2H^2}e^j, \ \ \ \  \omega_{ij}=\frac{f_{ij}}{2H^2}e^0+H^{-\frac{3}{2}}\delta_{k[i}\partial_{j]}H
e^k},
\\\\ \displaystyle{F=\frac{\sqrt{3}}{2H^2}\left[-H^{\frac{1}{2}}\partial_i H e^i\wedge e^0+\frac{f_{ij}}{2}e^i\wedge e^j\right], \ \ \ \ d=e^0 H\partial_t +H^{-\frac{1}{2}}e^i\left(\partial_i-a_i\partial_t \right)}.
\earr
\label{spinc}
\eequ 
The non-trivial supersymmetry variation of the $D=5$ simple supergravity theory for our bosonic background is the gravitino variation\footnote{We follow the conventions of \cite{nicolai}. However, the signature choice therein is $(+----)$, so that our equations differ by a few factors of i. We choose for the flat Gamma matrices $\Gamma^{01234}=i$, and make use of the four dimensional Majorana representation plus $\Gamma^4=i\Gamma^{0}\Gamma^1\Gamma^2\Gamma^3$. Therefore, $\Gamma^0,\Gamma^1,\Gamma^2,\Gamma^3$ are real and $\Gamma^4$ purely imaginary. Furthermore, $\Gamma^0,\Gamma^4$ are antisymmetric and the remaining gamma matrices symmetric.}:

\bequ
\barr{l}
\displaystyle{\delta \Psi =d\epsilon +\frac{1}{4}\omega_{ab}\Gamma^{ab}\epsilon +\frac{i}{4\sqrt{3}}\left(e^a\Gamma^{bc}_{\ \ \ a}-4e^b\Gamma^c\right)F_{bc}\epsilon}
\\\\\ \ \ \ \  = \displaystyle{e^0\left(H\partial_t \epsilon +\left[\frac{f_{ij}\Gamma^{ij}}{8H^2}-\frac{i\partial_{i}H \Gamma^{i}}{2H^{\frac{3}{2}}}\right]\left(1-i\Gamma^{0}\right)\epsilon\right)+e^k\left(H^{-\frac{1}{2}}\left(\partial_k-a_k \partial_t\right)\epsilon +\frac{i\Gamma^{0}\partial_k H}{2H^{\frac{3}{2}}}\epsilon - \right.}
\\\\ \ \ \ \ \ \ \ \ - \displaystyle{\left. \left[\frac{\partial_i H \Gamma^{i}_{ \ k}}{4H^{\frac{3}{2}}}+\frac{if_{ki}\Gamma^{i}}{2H^2}\right]\left(1-i\Gamma^{0}\right)\epsilon -\frac{\Gamma^{i0}}{4H^2}\left(f_{ik}-\star f_{ik}\right)\epsilon\right).}
\earr
\eequ
We have introduced the notation $f_{ij}$ for the components of the two form $f=da$, and $\star f$ is the Hodge dual of $f$ on ${\Bbb E}^4$. One can check that the $f$ that follows from (\ref{ah}) (or its multi black hole generalisation) is a self dual form on ${\Bbb E}^4$. Then, for 
\bequ
\epsilon =H^{-\frac{1}{2}}\epsilon_0^{K}, \ \ \  \epsilon_0^{K}=i\Gamma^0 \epsilon_0^{K},
\eequ
we get $\delta \Psi =0$, i.e. such $\epsilon$ is a Killing spinor. However, if one chooses the constant spinor $\epsilon_0$ obeying
\bequ
\epsilon_0^{AK}=-i\Gamma^0 \epsilon_0^{AK},
\label{antikill}
\eequ
we get a non-trivial gravitino variation, which can be fed back into the vielbein and gauge field variations to yield the first order superpartners of the BMPV spacetime. These spinors are often called `anti-Killing' spinors. We will comment more on the choice of this specific form for the `anti-Killing' spinors below.

We now generate the superpartners for the BMPV (multi)-black hole spacetimes. To first non-trivial order the variations of the gravitino and gauge field are:
\bequ
\barr{l}
\displaystyle{\delta \Psi =-\frac{e^0}{H^2}\left[iH^{\frac{1}{2}}\partial_i H \Gamma^i-\frac{1}{4}f_{ij}\Gamma^{ij}\right]\epsilon -\frac{e^k}{H^{\frac{3}{2}}}\left[\partial_k H+\frac{1}{2}\partial_i H \Gamma^i_{ \  k}+\frac{if_{ki}\Gamma^i}{H^{\frac{1}{2}}}\right]\epsilon},
\\\\ \displaystyle{\delta A=\frac{\sqrt{3}i}{2}\left(\bar{\epsilon}\Psi -\bar{\Psi} \epsilon\right)=\frac{\sqrt{3}i}{2H^2}\left[\frac{e^0 f_i^{\ k}}{2}-H^{\frac{1}{2}}e^k\partial_i H\right]\left(\bar{\epsilon}\Gamma^i_{\ k}\epsilon\right)},
\earr
\eequ
while for the vielbein they follow from $\delta e^{a}=\bar{\epsilon}\Gamma^{a}\Psi-\bar{\Psi}\Gamma^a \epsilon$ yielding
\bequ
\barr{l}
\displaystyle{\delta e^{0}=\frac{i}{H^2}\left[\frac{e^0}{2}f_i^{\ k}-H^{\frac{1}{2}}e^k\partial_i H\right]\left(\bar{\epsilon}\Gamma^i_{\ k}\epsilon\right)},
\ \  \displaystyle{\delta e^{j}=-2iH^{-2}\left[f_{ik}e^k-e^0 H^{\frac{1}{2}}\partial_i H\right]\left(\bar{\epsilon}\Gamma^{ij}\epsilon\right)}.
\earr
\eequ
In the holonomic basis the first order superpartner takes the following form:
\bequ
\barr{l}
\displaystyle{ds^2=-H^{-2}\left[1+iH^{-2}f_i^{\ j}\left(\bar{\epsilon}\Gamma^i_{\ j}\epsilon\right)\right]dt^2-2H^{-2}\left[a_k+i\left(H^{-2}a_k f_i^{\ j}-3\partial_i H \delta^j_k\right)\left(\bar{\epsilon}\Gamma^i_{\ j}\epsilon\right)\right]dt dx^k}
\\\\ \ \ \ \ \ \ \ \ \ \displaystyle{+H\left[dx^k dx_k +\frac{4i}{H}\left(H^{-1}\partial_i H a_k +f_{ki}\right)\left(\bar{\epsilon}\Gamma^i_{\ j}\epsilon\right)dx^k dx^j\right]},
\earr
\eequ
\bequ
A=\frac{\sqrt{3}}{2H}\left[1+\frac{i}{2H^{2}}f_i^{\ j}\left(\bar{\epsilon}\Gamma^i_{\ j}\epsilon\right)\right]dt+\frac{\sqrt{3}}{2H}\left[a_k+i\left(\frac{1}{2H^{2}}a_k f_i^{\ j}-\partial_i H \delta^j_k\right)\left(\bar{\epsilon}\Gamma^i_{\ j}\epsilon\right)\right]dx^k.
\eequ
Comparing the three terms of $g_{0i}$ with the three terms of $A_i$, we see that they have similar form, but with different coefficients:
\begin{description}
\item[i)] The term with purely bosonic angular momentum, $a_k$, implies a relation  $J^{ki}=\mu^{ki}/\sqrt{3}$, leading to a gyromagnetic ratio $g=3$ as seen before;
\item[ii)] The term with purely fermionic angular momentum, $\partial H \epsilon^2$, gives $J^{ki}=\sqrt{3}\mu^{ki}$, implying $g=1$. This should be regarded as the gyromagnetic ratio for the fermionic superpartners of a static background;
\item[iii)] The `mixed' term, $a f \epsilon^2$, yields $J^{ki}=2/\sqrt{3} \mu^{ki}$ and gives rise to $g=3/2$.
\end{description}
As we move along the supermultiplet, the gyromagnetic ratio varies. By this we mean that $g$ for the BMPV black hole differs from the one for its superpartners. But notice that the supermultiplets are labelled by some value of the bosonic angular momentum, so that the superpartners of the $D=5$ RN black hole are not in the same supermultiplet as the BMPV black hole with $a \neq 0$. 

The result to keep in mind is that the gyromagnetic ratio for the $5D$ black holes with some kind of angular momentum seems to cover the range $1\le g\le 3$, with the extreme values attained for purely fermionic and bosonic angular momentum respectively. A similar behaviour will be seen when we study the Dirac equation.

Let us now justify the use of the form (\ref{antikill}) for the anti-Killing spinors. It is quite clear that any choice of $\epsilon=f(x)\epsilon_0$, where $f\ne H^{-\frac{1}{2}}$ or $\epsilon \ne \epsilon^{K}$, will lead to a non trivial gravitino variation and hence some kind of `superpartners'. The new set of fields will represent the same physical spacetime configuration as the initial one, just in a different `superframe'. This is in a very direct analogy with the excitation of the magnetic field by going to a frame moving with respect to a purely electric source. Then, we still have the same physical setup, i.e. some configuration of electric sources; the fact that we see a different set of fields (namely electric and magnetic) results from the fact that only the components of the electric plus the components of the magnetic field fit into a complete `relativistic' multiplet, while each of these fields would fill a `Newtonian' multiplet. Of course, the fundamental point is that the Lorentz group is a more fundamental symmetry than the Galilean group, which is a low energy approximation. Similarly, the superdiffeomorphisms group of a supergravity theory allows the same physical setup to be described by many different sets of fields. In many physical situations some of these sets of fields contain fermionic excitations while others will be purely bosonic. We can therefore gauge away the fermions, which can then be labelled as pure gauge. In many situations, however, the fermions cannot be gauged away (In the context of $N=1,D=4$ Supergravity, necessary and sufficient conditions for an excited gravitino to be pure gauge were given in \cite{urban}).

One way to deal with such gauge arbitrariness is to impose the tracelessness condition to the first order gravitino, i.e., $\Gamma^{\mu}\Psi_{\mu}=0$ \cite{bko3}. Physically this means that the first order gravitino is a pure spin $3/2$ excitation, since the $\Gamma^{\mu}\Psi_{\mu}$ projection of the gravitino transforms as a spin $1/2$ representation of the Lorentz group universal covering. Moreover, since $\Psi_{\mu}=\hat{D}_{\mu} \epsilon$, where $\hat{D}_{\mu}$ is the supercovariant derivative, we impose a condition on the existence of superpartners, namely that the candidate background has regular solutions to a modified Dirac equation $\Gamma^{\mu}\hat{D}_{\mu}\epsilon=0$. Of course, we are interested in $\epsilon$ obeying this equation but with $\hat{D}_{\mu} \epsilon \neq 0$, in order to get non-trivial superpartners. Our choice for anti-Killing spinors satisfies this criterion.

It has also been argued that a more fundamental criterion for suitable superpartners is to require the first order gravitino to be normalisable \cite{win}. For the superpartners of the static background it is very easy to show this is the case, just as for the $D=4$ RN spacetime \cite{bko3}:
\bequ
|\Psi|^2=\int_{\Sigma}d^4x\sqrt{g_{(4)}}\Psi^{\dagger}_{\mu}\Psi_{\nu}g^{\mu \nu}=2\pi^2 M_{ADM}((\epsilon_0^{AK})^{\dagger}\epsilon_{0}^{AK}),
\eequ
where $\Sigma$ is a spacelike surface.

\subsubsection{The gyromagnetic ratio for an electron}
A test Dirac fermion with charge $q$ and mass $m$ interacting with gravitational and electromagnetic fields is described by a wave function obeying the minimally coupled Dirac equation, which takes the standard form
\bequ
\left[\Gamma^{\mu}\left(\partial_{\mu}+\frac{1}{4}\omega_{ab\mu}\Gamma^{ab}+iqA_{\mu}\right)-m\right]\Psi=0.
\eequ
Using (\ref{spinc}), the spin connection term can be rewritten in terms of the Maxwell tensor:
\bequ
\left[\Gamma^{\mu}\left(\partial_{\mu}+iqA_{\mu}\right)-\frac{1}{4\sqrt{3}}\Gamma^{0}F_{ab}\Gamma^{ab}-m\right]\Psi=0.
\label{diracnon}
\eequ
where $\Gamma^0$ is the flat gamma matrix. Hence we see that the problem of studying a Dirac fermion in the background (\ref{fields}) can be restated as the study of a non-minimally coupled Dirac fermion interacting solely with an electromagnetic field. Of course this should be expected for a supersymmetric background, since the vanishing of the supercovariant derivative acting on some spinor means that there will be a cancellation of terms between the Maxwell field and the spin connection. 

In four dimensional flat space, adding a non-minimal electromagnetic interaction of the type
\bequ
\frac{iq}{8m}\Delta g \Gamma^{\mu \nu}F_{\mu \nu},
\eequ
to the minimally coupled Dirac equation yields in the non relativistic limit, a magnetic dipole interaction in the Hamiltonian of the form
\bequ
H_{dipole}=-\frac{q}{2m}{\bf s} \cdot {\bf B}\left(2+\Delta g\right)
\eequ
The natural higher dimensional generalisation of this result yields for the gyromagnetic ratio associated with a Dirac fermion obeying (\ref{diracnon}) the non-standard value $g=2+\Delta g$ and
\bequ
\Delta g \Psi = \frac{2m}{\sqrt{3} q}i\Gamma^{0} \Psi
\eequ
Therefore we arrive at the interesting conclusion that the gyromagnetic ratio of a spin $1/2$ particle depends on its spinor direction being parallel to the anti-Killing or Killing spinors. In particular we should get $g=3$ for the latter case and $g=1$ for the former, when the Dirac fermion is BPS. It is clear that there is a `conspiracy' between the behaviour of the elementary particle and the behaviour of the black holes seen in the last section. Its meaning, however, is not quite clear. Moreover, this seems to be a particular property of the five dimensional family of black holes. For the  $D=4$ extreme KN, such conspiracy does not arise: $g=2$ both for the bosonic background \cite{carter} and for the superpartners of the extreme RN background \cite{aiche}. The Dirac equation in the extreme KN background cannot be easily cast into the non-minimally coupled flat space form. But a result similar to the five dimensional one would require $\Delta g=0$, i.e., the vanishing of the non-minimal coupling. This is not expected to occur. Indeed, even for the $D=4$ extreme RN, the Dirac equation can easily be put in a form similar to (\ref{diracnon}) with $\Delta g= 1/2$.

Let us close this section by discussing the possible String Theory counterpart of these results. Massive string states have a Schwarzchild radius greater than their Compton wave length; in other words their mass is greater than the Planck mass (in ten dimensions). This led to the suggestion that such states should be identified with some extremal black holes in supergravity \cite{malcape}. One possible check on this conjecture was made for gyromagnetic ratios. On the String Theory side, the gyromagnetic ratio for heterotic states in the presence of a background gauge field were computed in \cite{russo}. A matching was verified with the $g$ value for some black holes of $D=4$, $N=4$ supergravity coupled to 22 vector multiplets (i.e. the low energy field theory for heterotic on $T^6$) \cite{dufflr}. On the string theory side, the computation follows from the knowledge of the action for the heterotic string in the presence of the background field. In our case, the black hole does not correspond to string states but rather to D-brane states. Since the open strings describing the D-branes do not couple to the Ramond-Ramond charge it is not clear how one could compute $g$ for the microscopic configuration.

\section{Conclusions}
In this paper we argued that within all known BPS, rotating, asymptotically flat stringy black holes, the five dimensional case is rather special. And that one may use these special spacetimes to learn more about the connections between microscopic and macroscopic gravity. Our framework was toroidally compactified string theory, but one may embed the BMPV geometry in M-theory compactified on more general Calabi-Yau spaces \cite{sabra}. In section 2 we performed a comparison with the typical irregular case: the four dimensional extreme Kerr-Newman. The special properties arise not only from the Chern-Simons term \cite{GMTown} but also from the possibility of having a Hodge self dual rotation two-form. This is illustrated by studying a special class of gravitational waves. 

The rich causal structure of these spacetimes also presents a novel feature: CTC's homotopic to a point in the five dimensional spacetime are resolved in the universal covering of the ten dimensional uplifted geometry. Although CTC's are clearly a non-perturbative effect in string theory (as the repulson effect of section 4.1), they manifest themselves at weak coupling by the loss of unitarity. The point here is that states violating (\ref{unibou}) would imply the existence of states with negative $\tilde{L}_0$ eigenvalue. Consequently there would be negative norm states, i.e., non-unitary or ghost states.

Much controversy has surrounded causality as a fundamental principle in theories of gravity. Despite the doubts cast by the information paradox, unitarity is generally a more solid principle in quantum theories. In the same way we still have to understand quantitatively the microscopic description of the most fundamental black hole - Schwarzchild - it will be an interesting problem to understand the microscopic states associated with more fundamental acausal spacetimes, as the G\"{o}del manifold, and scrutinise the role of unitarity.

Another interesting property of these black holes concerns the gyromagnetic ratio. In quantum field theory, requiring a `good' behaviour for the tree level scattering amplitudes singles out $g=2$ as the most natural value for elementary particles \cite{ferpor}. This is implemented by a non-minimal electromagnetic coupling for fields with spin higher than $1/2$. For spin 1 charged matter, another argument in favour of $g=2$ is that a coupling consistent with such gyromagnetic ratio gives rise to a non electromagnetic gauge symmetry \cite{jackiw}. One could argue that a similar statement applies to black holes, if one is trying to interpret them as field theory realizations of fundamental states. It would therefore be interesting to understand in the microscopic context  the results of section 4.2.

\section *{Acknowledgments}
I would like to thank Miguel Costa, Malcolm Perry, Harvey Reall and Paul Townsend for discussions. I am particularly grateful to Gary Gibbons for many discussions and suggestions. The author is supported by FCT (Portugal) through grant no. PRAXIS XXI/BD/13384/97. This work is also supported by the PPARC grant PPA/G/S/1998/00613.


\newpage
\begin{thebibliography}{40}

\bibitem{cvey4}M.Cveti\v{c}, D.Youm, {\em Entropy of Non-Extreme Charged Rotating Black Holes in String Theory}, Phys. Rev. {\bf D54} (1996) 2612, hep-th/9603147.
\bibitem{cveyoum}M.Cveti\v{c}, D.Youm, {\em General rotating five-dimensional black holes of toroidally compactified heterotic string}, hep-th/9603100, Nucl. Phys. {\bf B476} (1996) 118.
\bibitem{cvey6}M.Cveti\v{c}, D.Youm, {\em Near-BPS-Saturated Rotating Electrically Charged Black Holes as String States}, Nucl. Phys. {\bf B477} (1996) 449, hep-th/9605051.
\bibitem{BMPV}J.Breckenridge, R.Myers, A.Peet, C.Vafa, {\em D-branes and Spinning Black Holes}, hep-th/9602065, Phys. Lett. {\bf B391} (1997) 93.
\bibitem{GMTown}J.Gauntlett, R.Myers, P.Townsend, {\em Black Holes of D=5 Supergravity}, hep-th/9810204, Class. Quant. Grav. {\bf 16} (1999) 1.
\bibitem{cremmer}E.Cremmer, {\em Supergravities in 5 dimensions}, in ``Superspace and Supergravity'', ed. S.W.Hawking, M.Ro\v{c}ek, Cambridge University Press  (1980) 267.
\bibitem{cve11}M.Cveti\v{c}, D.Youm, {\em Rotating Intersecting M-branes} Nucl. Phys. {\bf B499} (1997) 253, hep-th/9612229.
\bibitem{maldaphd}J.Maldacena, {\em Black Holes in String Theory}, Ph.D. thesis, hep-th/9607235.
\bibitem{bko1}E.Berghoeff, R.Kallosh, T.Ortin, {\em Supersymmetric String Waves}, hep-th/9212030, Phys. Rev. {\bf D47} (1993) 5444.
\bibitem{carter}B.Carter, {\em Global Structure of the Kerr family of Gravitational Fields}, Phys. Rev. {\bf 174} (1968) 1559.
\bibitem{hawell}S.Hawking, G.Ellis, {\em The large scale structure of the space-time}, Cambridge University Press (1973).


\bibitem{gibher}G.Gibbons, C.Herdeiro, {\em Supersymmetric Rotating Black Holes and Causality Violation}, hep-th/9906098, Class. Quant. Grav. {\bf 16} (1999) 3619. 
\bibitem{sen}A.Sen, {\em Rotating Charged Black Hole solution in Heterotic String Theory}, Phys. Rev. Lett. {\bf 69} (1992) 1006.
\bibitem{dufflr}M.Duff, J.Liu, J.Rahmfeld, {\em Dipole Moments of Black Holes and String States}, Nucl. Phys. {\bf B494} (1997) 161, hep-th/9612015.
\bibitem{balaktw}V.Balasubramanian, D.Kastor, J.Traschen, K.Win, {\em The Spin of the M2-Brane and Spin-Spin Interactions via Probe Techniques}, hep-th/9811037, Phys.Rev. {\bf D59} (1999) 084007.
\bibitem{larsen}F.Larsen, {\em Rotating Kaluza-Klein black holes}, hep-th/9909102.
\bibitem{jap}A.Hosoya, K.Ishikawa, Y.Ohkuwa, K.Yamagishi, {\em Gyromagnetic ratio of heavy particles in Kaluza-Klein theory}, Phys. Lett. {\bf 134B} (1984) 44.
\bibitem{duffg1}M.Duff, J.Liu, J.Rahmfeld, {\em g=1 for Dirichlet 0-branes}, hep-th/9810072, Nucl. Phys. {\bf B524} (1998) 129.
\bibitem{witten}E.Witten, {\em String Theory Dynamics In Various Dimensions}, Nucl. Phys. {\bf B443} (1995) 85, hep-th/9503124. 
\bibitem{aiche}P.Aichelburg, F.Embacher, {\em Exact superpartners of $N=2$ supergravity solitons}, Phys. Rev. {\bf D34} (1986) 3006.
\bibitem{clamof}M.Clayton, J.Moffat, {\em Scalar-tensor gravity theory for dynamical light velocity}, gr-qc/9910112.


\bibitem{calmal}C.Callan, J.Maldacena, {\em D-brane approach to black hole quantum mechanics}, Nucl. Phys. {\bf B472} (1996) 591; hep-th/9602043.
\bibitem{traschen}G.Gibbons, D.Kastor, L.London, P.Townsend, J.Traschen, {\em Supersymmetric self-gravitating solitons}, Nucl. Phys. {\bf B416} (1994) 850.
\bibitem{godel}K.G\"{o}del, {\em An example of a new type of cosmological solutions of Einstein's field equations of gravitation}, Rev. Mod. Phys {\bf 21} (1949) 447.
\bibitem{accgr}S.Chakrabarti, R.Geroch, C. Liang, {\em Timelike curves of limited acceleration in general relativity}, Jour. Math. Phys. {\bf 24} (1983) 597.
\bibitem{horw}G.Horowitz, S.Ross, {\em Naked Black Holes}, hep-th/9704058, Phys. Rev. {\bf D56} (1997) 2180.
\bibitem{nicolai}A.Chamaseddine, H.Nicolai, {\em Coupling the SO(2) Supergravity theory through dimensional reduction}, Phys. Lett. {\bf 96B} (1980) 89.
\bibitem{urban}P.Aichelburg, H.Urbantke, {\em Necessary and Sufficient Conditions for trivial solutions in Supergravity}, Gener. Rel. Grav. {\bf 13} (1981) 817.
\bibitem{gibgauge}G.Gibbons in Supersymmetry, Supergravity and related topics, Proceedings of the XVth GIFT seminar, ed. F.Aguila et al., World Scientific, Singapore, 1985.
\bibitem{bko3}R.Brooks, R.Kallosh, T.Ort\'{i}n, {\em Fermionic zero modes and black hole hypermultiplet with Rigid Supersymmetry}, hep-th/9505116.
\bibitem{win}K.Win, {\em Fermion Zero modes and Cosmological Constant}, hep-th/9811082.
\bibitem{celes}C.Duval, G.Gibbons, P.Horv\'{a}thy, {\em Celestial mechanics, conformal structures, and gravitational waves}, Phys. Rev. {\bf D43} (1991) 3907.
\bibitem{duffss}M.Duff, R.Khuri, J.Lu, {\em String Solitons}, Phys. Rept. {\bf 259} (1995) 213; hep-th/9412184. 
\bibitem{vafa}C.Vafa, {\em Gas of D-branes and the Hagedorn Density of BPS states}, Nucl. Phys. {\bf B463} (1996) 415; hep-th/9511088.
\bibitem{kent}W.Bucher, D. Friedan, A.Kent, {\em Determinant Formulae and Unitarity for the N=2 Superconformal algebras in two dimensions, or exact results in string compactification}, Phys. Lett. {\bf B172} (1986) 316.
\bibitem{peetpol}C.Johnson, A.Peet, J.Polchinski, {\em Gauge Theory and the Excision of Repulson Singularities}, hep-th/9911161.
\bibitem{cardy}J.Cardy, {\em Operator Content of two dimensional conformally invariant theories}, Nucl. Phys. {\bf B270} (1986) 186; see also J.Polchinski {\em String Theory}, Cambridge University Press (1998), chapters 7 and 14.



\bibitem{tod1}K.Tod, {\em All metrics admitting super-covariantly constant spinors}, Phys. Lett. {\bf 121B} (1983) 241.
\bibitem{iwp}W.Israel, G.Wilson, {\em A class of Stationary Electromagnetic Vacuum Fields}, J. Math. Phys. {\bf 13} (1972) 865; Z.Perj\'{e}s, {\em Solutions of the Coupled Einstein-Maxwell Equations representing the fields of Spinning sources}, Phys. Rev. Lett. {\bf 27} (1971) 1668. 
\bibitem{harhaw}J.Hartle, S.Hawking, {\em Solutions of the Einstein-Maxwell Equations with Many Black Holes}, Comm. Math. Phys. {\bf 26} (1972) 87.
\bibitem{pol}J.Polchinski, {\em String Theory}, Cambridge University Press, (1998), chapter 2.
\bibitem{ransun}L.Randall, R.Sundrum, {\em An Alternative to Compactification}, Phys. Rev. Lett. {\bf 83} (1999) 4690.
\bibitem{malcape}C.Callan, J.Maldacena, A.Peet, {\em Extremal Black Holes as fundamental strings}, Nucl. Phys. {\bf B475} (1996); hep-th/9510134.
\bibitem{russo}J.Russo, L.Susskind, {\em Asymptotic level density in heterotic string theory and rotating black holes}, Nucl. Phys. {\bf B437} (1995) 611.
\bibitem{york}J.York, {\em Gravitational degrees of freedom and the Initial Value problem}, Phys. Rev. Lett. {\bf 26} (1971) 1656.


\bibitem{ferpor}S.Ferrara, M.Porrati, V.Telegdi, {\em g=2 as the natural value of the tree level gyromagnetic ratio of elementary particles}, Phys. Rev. {\bf D46} (1992) 3529.
\bibitem{jackiw}R.Jackiw, {\em g=2 as a gauge condition}, Phys. Rev. {\bf D57} (1998) 2635, hep-th/9708097.
\bibitem{eisen}L.Eisenhart, {\em Riemannian Geometry}, Princeton University Press, 8th Printing (1997), p. 92.
\bibitem{papa}G. Papadopoulos, {\em Rotating Rotated branes}, JHEP {\bf 9904} (1999) 014, hep-th/9902166.
\bibitem{sabra}A.Chamseddine, W.Sabra, {\em Metrics Admitting Killing spinors in five dimensions}, Phys. Lett. {\bf B426} (1998) 36, hep-th/9801161.

\end{thebibliography}
\end{document}